\documentclass{aa}  

\usepackage{graphicx}
\usepackage{natbib}
\usepackage{placeins}
\usepackage{txfonts}
\usepackage{hyperref}
\hypersetup{
    colorlinks=true,
    linkcolor=blue,
    citecolor=blue,
    urlcolor=cyan,
    }

\usepackage{bm}
\usepackage{pdflscape}
\usepackage{rotating}
\usepackage{xcolor}
\usepackage{colortbl}

\usepackage{booktabs}
\defcitealias{Brun+2022}{B22}
\defcitealias{2019ApJS..244...21S}{S19}         
\defcitealias{santosSurfaceRotationPhotometric2021}{21}
\defcitealias{amardFirstGridsLowmass2019}{A19}  
\defcitealias{bergerGaiaKeplerStellarProperties2020}{B20}
\usepackage{amstext}

\begin{document}

   \title{Hunting for anti-solar differentially rotating stars using the Rossby number}

   \subtitle{An application to the \textit{Kepler} field}

   \author{Q. Noraz\inst{1}\thanks{E-mail: quentin.noraz@cea.fr}, S. N. Breton\inst{1}, A. S. Brun\inst{1}, R. A. Garc\'ia \inst{1}, A. Strugarek\inst{1}, A. R. G. Santos\inst{2,3}, S. Mathur\inst{4,5} \and L. Amard\inst{1}
          }

   \institute{D\'epartement d'Astrophysique/AIM, CEA/IRFU, CNRS/INSU, Univ. Paris-Saclay \& Univ. de Paris, 91191 Gif-sur-Yvette, France\\
             \and
             Instituto de Astrof\'isica e Ci\^encias do Espa\c{c}o, Universidade do Porto, CAUP, Rua das Estrelas, PT4150-762 Porto, Portugal \\
             \and
             Department of Physics, University of Warwick, Coventry, CV4 7AL, UK\\
             \and
             Instituto de Astrof\'isica de Canarias (IAC), E-38205 La Laguna, Tenerife, Spain\\
             \and
             Universidad de La Laguna (ULL), Departamento de Astrof\'isica, E-38206 La Laguna, Tenerife, Spain\\
             }

   \date{Received 28 April, 2022, Accepted 5 July, 2022}
 
  \abstract
   {Anti-solar differential rotation profiles have been found for decades in numerical simulations of convective envelopes of solar-type stars. These profiles are characterized by a slow equator and fast poles (i.e., reversed with respect to the Sun) and have been found in simulations for high Rossby numbers. Rotation profiles like this have been reported in evolved stars, but have never been unambiguously observed for cool solar-type stars on the main sequence. As solar-type stars age and spin down, their Rossby numbers increase, which could therefore induce a transition toward an anti-solar differential rotation regime before the end of the main sequence. Such a rotational transition will impact the large-scale dynamo process and the magnetic activity. In this context, detecting this regime in old main-sequence solar-type stars would improve our understanding of their magnetorotational evolution.}
   {The goal of this study is to identify the most promising cool main-sequence stellar candidates for anti-solar differential rotation in the \textit{Kepler} sample.}
   {First, we introduce a new theoretical formula to estimate fluid Rossby numbers, $Ro_{\rm f}$, of main-sequence solar-type stars. We derived it from observational quantities such as $T_{\rm eff}$ and $P_{\rm rot}$, and took the influence of the internal structure into account. Then, we applied it on a subset of the most recent catalog of \textit{Kepler} rotation periods, after removing subgiants and selecting targets with solar metallicity. Next, we considered the highest computed $Ro_{\rm f}$ and inspected each target individually to select the most reliable anti-solar candidate. Finally, we extended our study to stars with metallicities different from that of the Sun. To this end, we developed a formulation for $Ro_{\rm f}$ dependent on the metallicity index [Fe/H] by using 1D stellar grids, and we also considered this compositional aspect for the selection of the targets.}
   {We obtain a list of the most promising stars that are likely to show anti-solar differential rotation. We identify two samples: one at solar metallicity, including 14 targets, and another for other metallicities, including 8 targets. We find that the targets with the highest $Ro_{\rm f}$ are likely to be early-G or late-F stars at about log$_{10}g=4.37$~dex.}
   {We conclude that cool main-sequence stellar candidates for anti-solar differential rotation exist in the \textit{Kepler} sample. The most promising candidate is KIC~10907436, and two other particularly interesting candidates are the solar analog KIC~7189915 and the seismic target KIC~12117868. Future characterization of these 22 stars is expected to help us understand how dynamics can impact magnetic and rotational evolution of old solar-type stars at high Rossby number.}

   \keywords{Stars: rotation --
                Stars: fundamental parameters --
                Stars: evolution --
                Stars: solar-type --
                Sun: rotation --
                Sun: evolution --
                Techniques: photometric --
                Methods: observational --
                Methods: data analysis --
                Methods: analytical --
                Convection}
    
    \titlerunning{Searching for anti-solar DR in the \textit{Kepler} field}
    \authorrunning{Q. Noraz et al.}
   
   \maketitle
   
\section{Introduction}\label{sec:1}
Observations of stellar rotation show that solar-type stars experience various phases of rotational evolution during their life \citep{galletImprovedAngularMomentum2013}. In particular, their angular momentum decreases during the main-sequence (MS) phase, likely due to mass loss associated with their magnetized stellar wind (see \citealt{1962AnAp...25...18S}, \citealt{WeberDavis1967}, \citealt{mestelMagneticBrakingStellar1968} or \citealt{vidottoEvolutionSolarWind2021} for a recent review). The resulting spin-down is observationally well described by the Skumanich law $\Omega_* \propto t^{1/2}$ \citep{1972ApJ...171..565S} and enables estimating MS stellar ages from gyrochronology \citep{barnesRotationalEvolutionSolar2003}, within a given parameter space.

The evolution of the angular velocity $\Omega_*$ can strongly affect the internal dynamics of stars. In particular, numerical simulations show that the differential rotation (DR) profile changes as the star spins down (\citealt{brunDifferentialRotationOvershooting2017}, \citeyear{Brun+2022}). Fast-rotating models tend to exhibit Jupiter-like, cylindrical, or quenched rotation profiles, intermediately rotating models exhibit solar-type DR (fast equator, slow poles), and slow-rotating models switch to an anti-solar DR regime, in which surface rotation exhibits fast poles along with a slow equator.

More generally, these DR transitions are well characterized by the dimensionless Rossby number $Ro$, which quantifies the amplitude of the advective transport over the Coriolis force. These different DR regimes are expected to significantly influence the internal dynamics of the star and its evolution. \cite{norazImpactAntisolarDifferential2022} and \cite{Brun+2022} (hereafter \citetalias{Brun+2022}) showed that simulations of solar-type dynamos change in nature depending on the rotational regime of their convective envelope. Low $Ro$ models (fast rotators) tend to exhibit short (yearly) magnetic cycles, and intermediate $Ro$ models tend to exhibit long (decadal) cycles. Finally, high $Ro$ models (slow rotators) switching to an anti-solar DR regime are likely to exhibit stationary dynamos by losing their periodic magnetic reversals.

In addition, with the recent increase in the number of stars with measured rotation periods, 
it has been observed that MS stars undergo epochs of weakened \citep{vansadersWeakenedMagneticBraking2016} or stalled braking \citep{curtisWhenStalledStars2020}, deviating from the classical Skumanich law. Thus, techniques such as gyrochronology, which rely on the direct relation between rotation and age, need further investigations. For the apparently weakened braking at late ages, different physical explanations have been proposed. The first is a change in the dynamo process of slowly rotating stars, as suggested by recent observations \citep{metcalfeLBTPEPSISpectropolarimetry2019}. It has been shown that the magnetic torque provided by stellar winds mostly depends on large-scale magnetic modes \citep{revilleEFFECTMAGNETICTOPOLOGY2015,finleyEffectCombinedMagnetic2018}. However, the observational characterization of stellar magnetic topologies is difficult \citep{seeNondipolarMagneticFields2019}, and recent studies of ab initio dynamo models do not report strong changes in large-scale magnetic topology that could support this scenario (\citealt{norazImpactAntisolarDifferential2022}, \citetalias{Brun+2022}). A change in coronal properties could also explain this transition, making the wind less efficient to extract angular momentum \citep{ofionnagainSolarWindTime2018}. However, we still lack observational and theoretical constraints on the stellar mass-loss rate of MS stars, which would confirm or refute this option. In this context, it is crucial to understand what exactly occurs for the angular momentum loss of slowly rotating stars.\\
Finally, the possibility of observational bias has also been proposed because slower rotators might lose their surface spots. This would make the detection of stellar rotational modulations relatively difficult with photometric techniques. However, \cite{hallWeakenedMagneticBraking2021} confirmed evidence of a weakened magnetic breaking by using asteroseismology, which makes the possibility of a photometric-observational bias less probable. Still, observations of late weakened magnetic braking are currently under debate as other studies failed to find it for their respective samples \citep{lorenzo-oliveiraConstrainingEvolutionStellar2019,nascimentojr.RotationSolarAnalogs2020}. At the moment, rotation rates of old MS solar-type stars are therefore uncertain. This will influence the likelihood of finding anti-solar rotators.

The anti-solar DR regime has been observed in various numerical simulations of convective rotating spherical shells for several decades \citep{1977GApFD...8...93G,mattConvectionDifferentialRotation2011,2011AN....332..883K,guerreroDIFFERENTIALROTATIONSOLARLIKE2013,gastineSolarlikeAntisolarDifferential2014,fanSIMULATIONCONVECTIVEDYNAMO2014,kapylaConfirmationBistableStellar2014,karakMagneticallyControlledStellar2015,simitevDYNAMOEFFECTSTRANSITION2015,mabuchiDIFFERENTIALROTATIONMAGNETIZED2015,brunDifferentialRotationOvershooting2017,warneckeDynamoCyclesGlobal2018,strugarekSensitivityMagneticCycles2018,vivianiStellarDynamosTransition2019,Brun+2022}. Motivated by numerical results, different observational methods have been used so far to detect stellar DR: photometry, asteroseismology, and spectroscopy imaging. Using the latter, robust anti-solar detections have been obtained with Doppler imaging spectroscopy for red giants (see, e.g., \citealt{strassmeierDopplerImagingStellar2003}, \citealt{kovariAntisolarDifferentialRotation2017} and references therein) and subgiant stars \citep{harutyunyanAntisolarDifferentialRotation2016}. However, no unambiguous detection of such a profile has been reported so far for solar-type stars on the MS. The starspot signature can be tracked in photometric light-curves in order to quantify the velocity gradient $\Delta\Omega$ along the latitudes. Using this approach, \cite{reinholdDiscriminatingSolarAntisolar2015} asserted their belief that anti-solar DR is possible in 13 \textit{Kepler} targets. However, \cite{santosStarspotSignatureLight2017} showed that this problem is likely degenerated in most cases, making a clear distinction of the DR regime uncertain without independent constraints. Using asteroseismic methods, \cite{benomarAsteroseismicDetectionLatitudinal2018} detected signatures that suggested anti-solar DR profiles in ten solar-type stars in the \textit{Kepler} field, but the confidence level was not high enough to confirm detection. Thus, robust detections of anti-solar profiles are still pending for MS solar-type stars, and will need the scrutiny of the most promising targets with highly resolved data.

It is then instructive to assess whether the Sun or solar analogs could experience a change toward the anti-solar DR regime before they reach the end of their MS life. The solar fluid Rossby number $Ro_{\rm f}$ (see Sections \ref{sec:RossbyDefs} and \ref{sec:2Formula}) is currently thought to be between 0.6 and 0.9 according to recent numerical simulations in \citetalias{Brun+2022}, and the rotational transition is expected to be above $Ro_{\rm f}=1$. If the Sun were to strictly follow the Skumanich law \citep{1972ApJ...171..565S}, its $Ro_{\rm f}$ would increase by a factor 1.48 by the end of the MS. It could then exceed unity enough for an anti-solar DR regime to be possible \citep{gastineSolarlikeAntisolarDifferential2014,brunDifferentialRotationOvershooting2017}. Old solar analogs could therefore exhibit anti-solar DR if this were detectable with current techniques. More generally, different effects can play a role for the $Ro_{\rm f}$ value that the star can reach for solar-type stars. On the one hand, the more massive a solar-type star, the more vigorous its convective motions. However, the highest-mass solar-type stars may never rotate slowly enough to enter the anti-solar DR regime during the MS (see Fig.~6 of \citealt{amardFirstGridsLowmass2019}). On the other hand, the lower the mass of a solar-type star, the longer its MS lifetime and the higher its spin-down amplitude. Nevertheless, lower-mass stars tend to have lower Rossby numbers because their convective transport is weaker than their rotational constraint. Therefore, an intermediate-mass sweet spot is expected at which solar-type stars might be in the right parameter regime to counterbalance these two opposing effects and ultimately reach $Ro_{\rm f}$ , favoring the anti-solar DR regime.

The goal of the present work is to search for these stars. The \textit{Kepler} mission \citep{borucki2010Sci...327..977B} has observed almost 200,000 targets during its four-year-long nominal mission. Tens of thousands of these stars exhibit photometric rotational modulation \citep[e.g.,][]{2014ApJS..211...24M,2019ApJS..244...21S,santosSurfaceRotationPhotometric2021}. Hence, the \textit{Kepler} survey is currently the best-suited dataset in order to search for anti-solar rotating stars.

We identify the most promising targets for anti-solar DR detection in the sample of MS cool stars of the \textit{Kepler} field. In Sect.~\ref{sec:2} we present an improved theoretical derivation of the fluid Rossby number $Ro_{\rm f}$ compared to \cite{brunDifferentialRotationOvershooting2017}, taking the internal structure of solar-type stars into account and extending the computation for the use of the effective temperature, $T_{\rm eff}$. Then, we apply this formula to the {\it Kepler} catalog of \cite{2019ApJS..244...21S}, (\citeyear{santosSurfaceRotationPhotometric2021}) (referred to as S19-21) in Sect.~\ref{sec:3} for solar-metallic MS targets and select the highest $Ro_{\rm f}$ values. Next, we inspect each selected target individually and build a list 
of promising candidates, which are discussed. After this, we extend this method to other metallicities in Sect.~\ref{sec:4}. We finally discuss the limitations of our method in Sect.~\ref{sec:5} and conclude in Sect.~\ref{sec:6} by highlighting the most promising anti-solar candidates for a future characterization of these targets.

\section{Differential rotation states and Rossby number}\label{sec:2}
\subsection{Differential rotation profiles from numerical simulations}\label{sec:DRprofiles}
\begin{figure}
    \centering
    \includegraphics[width=\linewidth]{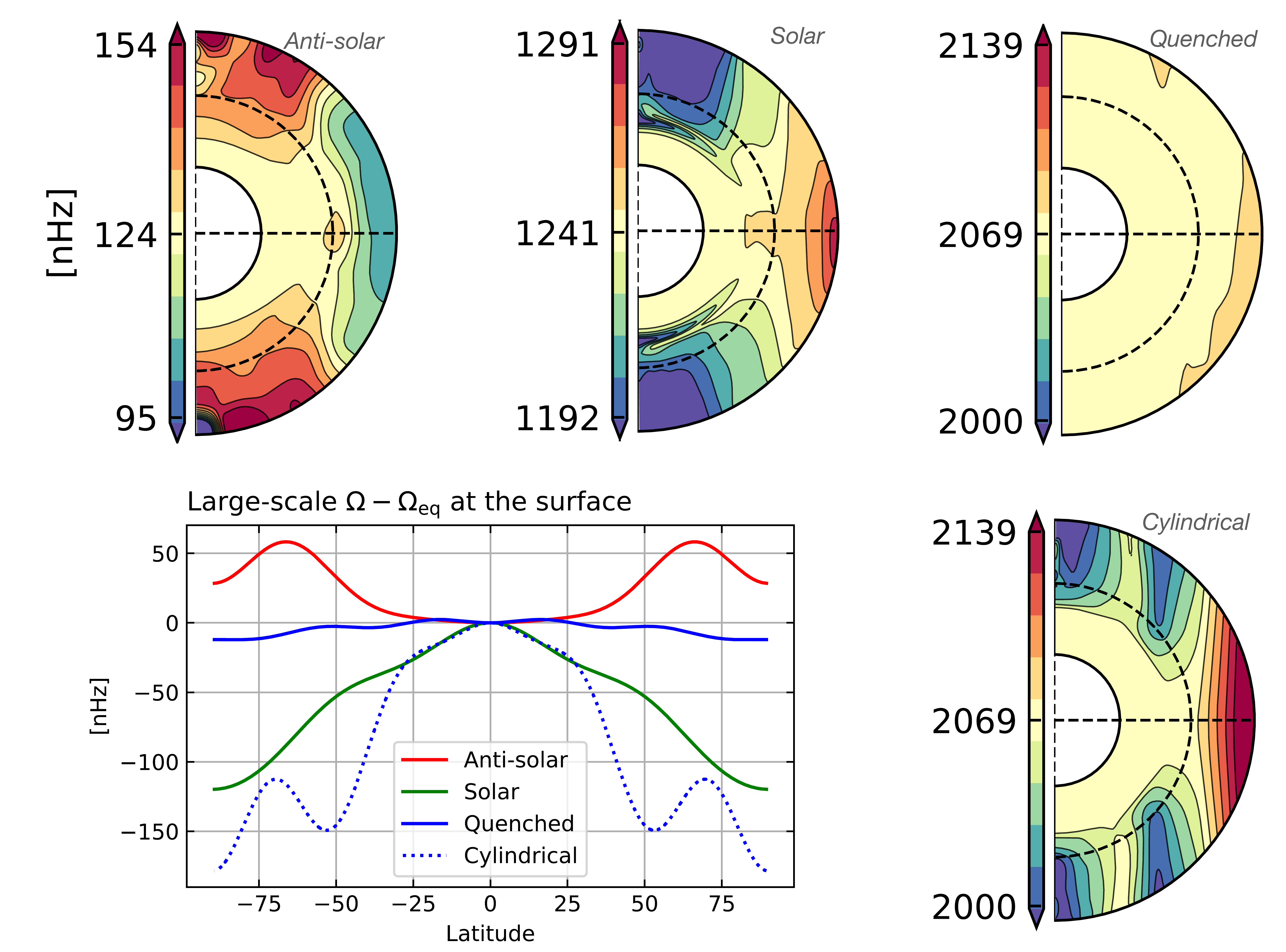}
    \caption{Example of DR profiles that emerge in numerical simulations of stellar convective envelopes. The top panels show DR profiles achieved in MHD models of \citetalias{Brun+2022}, from high to low Rossby numbers (left to right). The bottom right profile is characteristic of low Rossby numbers in hydrodynamics models of \cite{brunDifferentialRotationOvershooting2017}. The symmetric part of the surface velocity from these meridional cuts is illustrated in the bottom left panel for all profiles.}
    \label{fig:DRsummary}
\end{figure}

Numerical simulations of stellar convective envelopes yield various DR profiles, depending on the Rossby number (see \citealt{brunMagnetismDynamoAction2017} for a review). The dimensionless Rossby number $Ro$ parameterizes the influence of the rotation on the dynamics of a system. In a star, it quantifies the impact of the Coriolis force on turbulent convective motions. Convective motions can collectively transport angular momentum (through Reynolds stresses) to establish a large-scale DR. The characteristics of their DR therefore depend on $Ro$, as illustrated in Fig.~\ref{fig:DRsummary}. For low Rossby numbers (right panels), profiles are highly constrained by the influence of the rotation. In the bottom right panel, a hydrodynamical case shows a cylindrical profile, with alternating prograde and retrograde zonal jets (also called Jupiter-like jets). This alignment along the vertical axis results from the Taylor-Proudman constraint that was achieved in fast-rotating convective envelopes \citep{1975JAtS...32.1331G,2000ApJ...533..546E,mieschSolarDifferentialRotation2006}. The DR amplitude can be quenched when a magnetic field is considered to be in such a regime ($\Omega$-quenching) because of the Lorentz force feedback (\citealt{brunInteractionDifferentialRotation2004}, \citetalias{Brun+2022}). Around intermediate Sun-like Rossby numbers, typical solar and conical DR profiles are found, with fast equator and slow poles (top middle panel). Finally, anti-solar profiles with a slow equator and fast poles are found for high Rossby numbers (top left panel). The rotational transition from solar-type and anti-solar profiles lies around $Ro_{\rm f}=1$ \citep{gastineSolarlikeAntisolarDifferential2014}. We show in the bottom left panel the symmetric part of the surface DR that was achieved in these simulations. The amplitude of the shear of the anti-solar regime (about 50 nHz, i.e., 0.03 rad/days) is similar to that of the solar regime, which means that it might be detected with current capabilities (e.g., \citealt{reinholdRotationDifferentialRotation2015}).

\subsection{Rossby number definitions}\label{sec:RossbyDefs}
Different definitions of the Rossby number exist and are used in the community. While their absolute values can differ, they generally scale with each other (\citealt{landinTheoreticalValuesConvective2010}, \citetalias{Brun+2022}).\\
First, the stellar Rossby $Ro_{\rm s} = \frac{P_{\rm rot}}{\tau_{\rm conv}}$ is often used in the observational context. It compares two characteristic timescales, the rotation period of the star $P_{\rm rot}=2\pi/\Omega_*$ and its convective over-turning time $\tau_{\rm conv}$. Rotation can be measured directly (see, e.g., Sect.~4.1 of \citealt{brunMagnetismDynamoAction2017} and Sect.~7 of \citealt{garciaAsteroseismologySolartypeStars2019} for reviews), but $\tau_{\rm conv}$ has to be inferred from stellar models, considering mass, spectra, and chemical composition (see, e.g., \citealt{noyesRelationStellarRotation1984}, Fig.~6 of \citealt{cranmerTESTINGPREDICTIVETHEORETICAL2011}, or \citealt{corsaroCalibrationRossbyNumber2021}). Furthermore, $\tau_{\rm conv}$ can be computed at different depths in the star, which could also explain some differences in the absolute Rossby values.\\
An alternative is the fluid Rossby $Ro_{\rm f}=\omega/2\Omega_*=\omega P_{\rm rot}/4\pi$, which we chose to consider in this study, and where $\omega$ is the vorticity. It is often used to characterize numerical simulations because it quantifies a force balance between nonlinear advection and Coriolis acceleration in the Navier-Stokes equation. Once again, the vorticity $\omega$ is not observed and has to be deduced from models. Similarly to $Ro_{\rm s}$, this quantity depends on the stellar depth that is chosen. We therefore propose in the next subsection to infer it from observable quantities, and to normalize it to the solar value in order to alleviate a dependence on the $Ro$ definition we chose.

\subsection{Fluid Rossby number based on observable quantities}\label{sec:2Formula}
First, we derived an analytical formulation of $Ro_{\rm f}$ as a function of the stellar mass $M_*$ and the rotation rate $\Omega_*$, following \cite{brunDifferentialRotationOvershooting2017}. We then extended it by considering the effective temperature $T_{\rm eff}$ in order to infer $Ro_{\rm f}$ from observations. Next, we calibrated it with scaling laws from stellar evolution models and compared the resulting trend with 3D global numerical simulations of stellar convection. Finally, we took a structural aspect into account in order to extend the mass range on which we can apply the final formula. This is presented in Eq.~\ref{eq:RofTeffProt}.

\subsubsection{Fluid Rossby number as a function of $M_*$ and $P_{\rm rot}$}
Based on fluid dynamics, $Ro_{\rm f}$ can be estimated as
\begin{equation}
    Ro_{\rm f}=\omega/2\Omega_*\simeq\dfrac{v_{\rm conv}}{2\Omega_*D}
    \label{eq:RofDef}
,\end{equation}
where $v_{\rm conv}$ is the representative convective velocity participating in the Reynolds stresses, $D$ is a characteristic length scale, and $\Omega_*$ the rotation rate of the star. In general, $D$ is considered to be the stellar radius $R_*$ or the convection zone (CZ) thickness.\\
Following the mixing-length theory (MLT) argument in \cite{brunDifferentialRotationOvershooting2017}, the enthalpy flux $F_{\rm en}$ transported by convective cells is the main source of energy transport through the CZ of cool stars. We can relate $F_{\rm en}$ to the velocity by assuming that $F_{\rm en}\sim\rho v^3$, by analogy to the kinetic energy flux. Hence, we have $L_*\propto \rho_{BCZ}v_{\rm conv}^3R_*^2$, with $\rho_{BCZ}$ the density at the base of the convection zone (BCZ). The fluid Rossby number is then proportional to
\begin{equation}
    Ro_{\rm f}\propto \left(\dfrac{L_*}{\rho_{BCZ}R_*^5}\right)^{1/3}\Omega_*^{-1}.\end{equation}
We used the well-known homology relations on the MS \citep{kippenhahnStellarStructureEvolution2012}, such as
\begin{equation}
    L_*\propto M_*^m\hspace{0.5cm}\text{and}\hspace{0.5cm}R_*\propto M_*^n,
    \label{eq:ScalingLawsL&R}
\end{equation}
and we parameterized $\rho_{BCZ}$ during the MS, as $\rho_{BCZ}\propto M_*^p$, to obtain
\begin{equation}
    Ro_{\rm f}\propto M_*^{(m-5n-p)/3} \Omega_*^{-1},
    \label{eq:MLTRofM}
\end{equation}
with $m,n>0$ and $p<0,$ but undetermined for now. We obtained an expression of $Ro_{\rm f}$ based on the mass and rotation rate of a solar-type star. We now proceed and derive its equivalent formulation based on observable quantities.

\subsubsection{Observable quantities}
One main point of this paper is to infer the fluid Rossby number through the use of $T_{\rm eff}$ rather than $M_*$. The determination of $M_*$ is model dependent and therefore has larger uncertainties \citep{lebretonAsteroseismologyCarteStellar2014}. On the other hand, $T_{\rm eff}$ can be directly measured with high precision through spectroscopy (see, e.g., APOGEE DR17; \citealt{abdurroufSeventeenthDataRelease2022}).\\
Hence, we now consider the link between stellar luminosity $L_*$ and global quantities, such as the stellar radius $R_*$ and $T_{\rm eff}$,
\begin{equation}
    L_*=4\pi R_*^2\sigma T_{\rm eff}^4,\end{equation}
where $\sigma$ is the Stefan–Boltzmann constant. Using again the homology relations in Eq. \ref{eq:ScalingLawsL&R}, we then obtain the relation
\begin{equation}
    T_{\rm eff}\propto M_*^{(m-2n)/4},
    \label{eq:TeffM*}
\end{equation}
which subsequently gives
\begin{equation}
    Ro_{\rm f}\propto T_{\rm eff}^{\dfrac{4}{m-2n}\left(\dfrac{m-5n-p}{3}\right)} \Omega_*^{-1}.
    \label{eq:MLTRofT}
\end{equation}
We now have two analytical expressions of the $Ro_{\rm f}$ of a solar-type star, based on $\Omega_*$, along with $M_*$ or $T_{\rm eff}$ (Eqs.~\ref{eq:MLTRofM} and \ref{eq:MLTRofT}, respectively). We now compute the corresponding exponents based on various numerical models.

\subsubsection{Inferring scaling indexes from numerical models}\label{sec:NumModels}
In order to compute exponents of Eqs.~\ref{eq:MLTRofM} and \ref{eq:MLTRofT}, we performed linear regression fits on the evolutionary tracks from \cite{amardFirstGridsLowmass2019} (hereafter \citetalias{amardFirstGridsLowmass2019}). We focused on solar-type stars and therefore considered masses ranging from $0.4$ to $1.3M_\odot$ and a metallicity index [Fe/H] from $-1$ to +0.3~dex. This yields $m=4.66\pm0.06$, $n=1.16\pm0.01$, and $p=-9.28\pm0.37$ for the MS. When applied to Eqs.~\ref{eq:MLTRofM} and \ref{eq:MLTRofT}, we obtain
\begin{equation}
    Ro_{\rm f}\propto M_*^{2.71\pm0.13}\Omega_*^{-1}\propto T_{\rm eff}^{4.64\pm0.23}\Omega_*^{-1}.
    \label{eq:MLTRofT_AN}
\end{equation}

For the sake of robustness, we now performed a multiparameter linear regression fits directly on the fluid Rossby numbers achieved by convective motions in the published 3D magnetohydrodynamcis (MHD) global simulations of \citetalias{Brun+2022}. We directly computed $Ro_{\rm f}=\omega/2\Omega_*$ from them, and assessed how it scales with $\Omega_*$, $M_*$ , and $T_{\rm eff}$ (see the tabulated $Ro_{\rm f}$ values in B22). B22 modeled solar-type stars for a mass range of $0.5<M_*<1.1$ M$_\odot$, a temperature range of $4030$\,K $<T_{\rm eff}<6030$\,K at different rotation rates $\Omega_*$ , and for the solar composition. This gives
\begin{equation}
    Ro_{\rm f}\propto\dfrac{M_*^{1.63\pm 0.24}}{\Omega_*^{0.9\pm 0.06}}\propto\dfrac{T_{\rm eff}^{3.2\pm0.4}}{\Omega_*^{0.9\pm 0.05}}.
    \label{eq:GKmagScales}
\end{equation}
We note that the exponents from Eq.~\ref{eq:MLTRofT_AN} computed from a broad sample of solar-type 1D models do not match with what is inferred from ab initio convective simulations in Eq.~\ref{eq:GKmagScales}. The stellar structure of solar-type stars can significantly change within a wide range of masses and metallicities, which subsequently impacts the $Ro_{\rm f}$ value. In particular, we note that 1.2 and 1.3 M$_\odot$ 1D models notably impact exponents inferred from linear fits. In the next section, we therefore add the structural aspect of solar-type stars in the analytical development of $Ro_{\rm f}$. The specific effect of metallicity is addressed in Section \ref{sec:4}.

\subsubsection{Structural fluid Rossby number}\label{sec:RofStruct}
In an attempt to take the stellar structure into account, the characteristic length $D$ (Eq.~\ref{eq:RofDef}) can be taken as the middle of the convective envelope $R_\text{mid}=0.5(R_{BCZ} + R_*)=0.5R_*(\alpha + 1),$ where $\alpha = R_{BCZ}/R_*$ is the aspect ratio of the radiative interior 
of the cool star. Following the derivation of the previous section, we find
\begin{equation}
    L_*\propto \bar{\rho}v_{\rm conv}^3R_{\rm mid}^2\propto\bar{\rho}v_{\rm conv}^3(1+\alpha)^2R_*^2\;,
    \label{eq:Lconv}
\end{equation}
with $\bar{\rho}$ the averaged density in the convective envelope. The combination of Eqs.~\ref{eq:RofDef} and \ref{eq:Lconv} gives
\begin{equation}
    Ro_{\rm f}\propto (1+\alpha)^{-5/3}\left(\dfrac{L_*}{\bar{\rho}R_*^5}\right)^{1/3}\Omega_*^{-1}.
    \label{eq:befDensityExpress}
\end{equation}
We now express the average density of the CZ as its mass divided by its volume,
\begin{equation}
    \bar{\rho}\sim\dfrac{M_{CZ}}{V_{CZ}}\propto\dfrac{(1-\beta)M_*}{(1-\alpha^3)R_*^3},
    \label{eq:rhoDecomp}
\end{equation}
where $\beta=M_{BCZ}/M_*$ is the relative mass of the radiative interior. 
Injecting it into Eq.~\ref{eq:befDensityExpress} leads to
\begin{equation}
    Ro_{\rm f}\propto S(\alpha,\beta)\left(\dfrac{L_*}{M_*R_*^2}\right)^{1/3}\Omega_*^{-1}
    \label{eq:RofSab}
,\end{equation}
\begin{equation}
    \text{where}\hspace{0.2cm} S(\alpha,\beta)=\left(\dfrac{1-\alpha^3}{(1-\beta)(1+\alpha)^5}\right)^{1/3}
    \label{eq:Sab}
\end{equation}
is a structure term depending on the structure parameters $\alpha$ and $\beta$. We finally assume that the stellar structure does not change during the MS, which only leaves a dependence on the stellar mass at first approximation. We can then parameterize it as $S(\alpha,\beta)\propto M_*^q$ during the MS, which yields
\begin{equation}
    Ro_{\rm f}\propto M_*^a\Omega_*^{-1}\;\;\text{where}\;\;a=\dfrac{m-(2n+1)}{3}+q,
    \label{eq:RofStructM*}
\end{equation}
with $q>0$ a priori, but undetermined for now. Using previous development, this gives
\begin{equation}
    Ro_{\rm f}\propto T_{\rm eff}^b\Omega_*^{-1}\;\;\text{where}\;\;b=\dfrac{4}{m-2n}\left(\dfrac{m-(2n+1)}{3}+q\right).
    \label{eq:RofStructTeff}
\end{equation}
Similarly to Sect.~\ref{sec:NumModels}, we computed the exponents of Eqs.~\ref{eq:RofStructM*} and \ref{eq:RofStructTeff} from the evolutionary tracks of \citetalias{amardFirstGridsLowmass2019}, considering $0.4\leq M_*\leq1.3M_\odot$. Linear regression fits yield $m=4.66\pm0.03$, $n=1.16\pm0.01$ and $q=1.48\pm0.05$ for the MS (see the fits and values in Fig.~\ref{fig:ScalingLaws}). Hence, applying these exponents to Eqs.~\ref{eq:RofStructM*} and \ref{eq:RofStructTeff}, we finally obtained the following relations:
\begin{equation}
    Ro_{\rm f}\propto\dfrac{M_*^{1.93\pm0.05}}{\Omega_*}\propto\dfrac{T_{\rm eff}^{3.29\pm0.09}}{\Omega_*}
    \label{eq:A19indexes}
.\end{equation}
We note that it is now closer to what we derived from numerical simulations in Eq.~\ref{eq:GKmagScales}, especially with $T_{\rm eff}$.

Considering T$_{\rm eff,\odot}=5772$ K \citep{prsaNOMINALVALUESSELECTED2016}, P$_{\rm rot,\odot}=27.3$ days, a convenient formula we used is
\begin{equation}
    \dfrac{Ro_{\rm f}}{\rm Ro_{\rm f,\odot}} = \left(\dfrac{P_{\rm rot,*}}{\rm P_{\rm rot,\odot}}\right)\times\left(\dfrac{T_{\rm eff}}{\rm T_{\rm eff,\odot}}\right)^{3.29}.
    \label{eq:RofTeffProt}
\end{equation}
 Using the solar values as reference prevents introducing a bias related to the dependence on the convention chosen for the $Ro$ definition (see Sect.~\ref{sec:RossbyDefs}). The scaling law in Eq.~\ref{eq:RofTeffProt} can then readily be applied to estimate the $Ro_{\rm f}$ of a star with respect to the Sun. We now estimate $Ro_{\rm f}$ of those \textit{Kepler}-field stars for which a rotation period has been measured.

\section{Application to \textit{Kepler} stars with solar metallicity}\label{sec:3}
In this section, we consider solar metallicity targets. Targets with other metallicities are addressed in Sect.~\ref{sec:4}.

\subsection{The Kepler rotation catalog}
The stellar surface rotation catalog compiled by \citetalias{2019ApJS..244...21S}-\citetalias{santosSurfaceRotationPhotometric2021} provides rotation periods and photometric magnetic activity proxies \citep{2010Sci...329.1032G,2014A&A...562A.124M} for 55,232 FGKM MS and subgiant \textit{Kepler} targets. It extends the 
catalog from \citet{2014ApJS..211...24M} with 24,182 new detections and populates the slow-rotator region for all spectral types. This is particularly important for this work. The \citetalias{2019ApJS..244...21S}-\citetalias{santosSurfaceRotationPhotometric2021} catalog was built by implementing the method described in \citet{2016MNRAS.456..119C,2017A&A...605A.111C}, which combines the outputs of the autocorrelation function and of a time-frequency analysis.
Machine-learning methods and visual inspections have been performed in order to ensure the reliability of the results \citep{2021A&A...647A.125B}.

The \citetalias{2019ApJS..244...21S}-\citetalias{santosSurfaceRotationPhotometric2021} catalog includes three subsamples. The first subsample (aliased I) corresponds to targets that are classified as solar-type stars in the \cite{mathurRevisedStellarProperties2017} and \cite{bergerGaiaKeplerStellarProperties2020} (hereafter \citetalias{bergerGaiaKeplerStellarProperties2020}) stellar property catalogs. Stars for which their properties disagree are gathered in the two last subsamples (aliased II and III). We decided here to consider only subsample I, which represents 50,656 targets. We also considered $M_*$ and $T_{\rm eff}$ proposed in \citetalias{2019ApJS..244...21S}-\citetalias{santosSurfaceRotationPhotometric2021}, which originated from \citetalias{bergerGaiaKeplerStellarProperties2020}.

\subsection{Data processing}\label{sec:PMS-SG}
First, in order to only select solar-metallicity stars, we extracted the metallicity index [Fe/H] for each star in the rotation catalog \citetalias{2019ApJS..244...21S}-\citetalias{santosSurfaceRotationPhotometric2021}. For the sake of accuracy, we took [Fe/H] from the APOGEE database (DR16: \citealt{ahumada16thDataRelease2020}, DR17: \citealt{abdurroufSeventeenthDataRelease2022}), LAMOST \citep{zhaoLAMOSTSpectralSurvey2012,catLAMOSTOBSERVATIONSKEPLER2015,zongLAMOSTObservationsKepler2018}, or \citetalias{bergerGaiaKeplerStellarProperties2020} in this order of priority for each target. We refer to Sect.~3.1 from \cite{mathurDetectionsSolarlikeOscillations2022} for more details. In this section, we consider stars at solar metallicity as stars with $-0.1$\,dex\,$<{\rm [Fe/H]}<0.1$\,dex. This means 16,258 stars of the 50,656 targets that were considered.

The criterion developed in Eq.~\ref{eq:RofTeffProt} applies only to MS stars. When it is blindly applied to pre-main-sequence (PMS) stars in the stellar evolution grid of \citetalias{amardFirstGridsLowmass2019}, we find a maximum $Ro_{\rm f}/{\rm Ro_{\rm f,\odot}}$ of 0.291, which is well below the anti-solar DR threshold above 1. Therefore, no explicit filtering of PMS stars is necessary because these stars are not expected to harbor anti-solar DR due to their fast rotation \citep{emeriau-viardOriginEvolutionMagnetic2017}. Because the relations are derived for MS stars and because of the way in which stars evolve past the terminal age main sequence (TAMS), we also excluded subgiants from our sample. Following \citetalias{2019ApJS..244...21S}-\citetalias{santosSurfaceRotationPhotometric2021}, we took the transition between MS and the subgiant branch from evolutionary tracks obtained from \citetalias{amardFirstGridsLowmass2019} (see Appendix \ref{sec:TAMSvsZ} for more details). Moreover, no subgiants for $M<0.8M_\odot$ up to $13.7$ Gyr are detected in these tracks. Hence, we assumed here that subgiants do not already exist at log$_{10}g>4.4$.

We also excluded stars that are flagged as type 1 classical pulsators/close binary (CP/CB) candidates in the \citetalias{2019ApJS..244...21S}-\citetalias{santosSurfaceRotationPhotometric2021} catalog. The characteristic period extracted from the light curves of these stars is likely to be distinct from the rotational behavior of single stars.

Finally, we only considered targets for which $0.4\leq M_*\leq1.3~M_\odot$ because this is the validity range for Eq.~\ref{eq:RofTeffProt}. The various data-processing steps presented in this section yielded a set of 12,517 stars. We then computed $Ro_{\rm f}$ for all stars in this sample.
 
\subsection{Most promising targets at solar metallicity}\label{sec:ZsunCandidates}

\begin{table*}
  \centering
  \caption{List and stellar parameters of anti-solar candidates at the solar metallicity}
  \label{tab:ZsunTargets}
  \begin{tabular}{lccccrclcc}
    \hline
    \hline\\ [-2.0ex]
    KIC &          log$_{10}$g &          $T_{\rm eff}$ [K] & $P_{\rm rot}$ [days] &  $Ro_{\rm f}/{\rm Ro_{\rm f,\odot}}$ &                  [Fe/H] &      $M_*$ [$M_\odot$] & $S_{\rm ph}$ [ppm] & Zfl & Sfl \\[0.5ex]
    \hline
    \hline\\ [-2.0ex]
    10907436 & $4.25_{-0.03}^{+0.04}$ & $6364.3_{-141.9}^{+137.1}$ &       $48.4 \pm 3.9$ &        $2.44_{-0.26}^{+0.26}$ &  $-0.07_{-0.03}^{+0.03}$ & $1.22_{-0.06}^{+0.06}$ &     $86.5 \pm 3.2$ &   1 &   2 \\[0.5ex] 
    8194513 & $4.40_{-0.04}^{+0.03}$ & $6030.1_{-119.4}^{+121.2}$ &       $44.5 \pm 3.6$ &        $1.88_{-0.20}^{+0.20}$ & $-0.05_{-0.16}^{+0.13}$ & $1.06_{-0.07}^{+0.06}$ &   $751.6 \pm 17.9$ &   2 &   1 \\[0.5ex] 
    6068129$^{+}$ & $4.40_{-0.04}^{+0.04}$ & $5838.9_{-110.8}^{+110.8}$ &       $49.4 \pm 8.6$ &        $1.88_{-0.35}^{+0.35}$ & $-0.01_{-0.16}^{+0.14}$ & $1.01_{-0.08}^{+0.07}$ &   $588.6 \pm 14.1$ &   2 &   1 \\[0.5ex] 
    12117868$^{*}$ & $4.24_{-0.04}^{+0.04}$ & $6068.0_{-114.8}^{+113.4}$ &       $42.5 \pm 5.0$ &        $1.84_{-0.24}^{+0.24}$ &  $-0.04_{-0.01}^{+0.01}$ & $1.09_{-0.07}^{+0.07}$ &     $71.7 \pm 1.9$ &   0 &   1 \\[0.5ex] 
    11912541 & $4.08_{-0.05}^{+0.04}$ & $5884.5_{-117.9}^{+113.9}$ &       $46.4 \pm 3.9$ &        $1.81_{-0.19}^{+0.19}$ &  $0.03_{-0.16}^{+0.14}$ & $1.13_{-0.09}^{+0.11}$ &   $419.0 \pm 10.9$ &   2 &   1 \\[0.5ex] 
    3746157 & $4.31_{-0.04}^{+0.04}$ & $5863.5_{-110.8}^{+115.2}$ &       $46.1 \pm 5.3$ &        $1.78_{-0.23}^{+0.23}$ &   $0.01_{-0.01}^{+0.01}$ & $1.03_{-0.07}^{+0.07}$ &    $446.7 \pm 9.8$ &   1 &   1 \\[0.5ex] 
    8245631 & $4.14_{-0.05}^{+0.05}$ & $5783.2_{-123.4}^{+124.0}$ &       $48.0 \pm 7.7$ &        $1.77_{-0.31}^{+0.31}$ &  $0.03_{-0.16}^{+0.17}$ & $1.05_{-0.08}^{+0.10}$ &   $380.2 \pm 11.5$ &   2 &   1 \\[0.5ex] 
    10670661 & $4.33_{-0.04}^{+0.04}$ & $5832.4_{-108.6}^{+111.8}$ &       $46.4 \pm 7.5$ &        $1.76_{-0.30}^{+0.30}$ &  $0.00_{-0.15}^{+0.16}$ & $1.01_{-0.08}^{+0.08}$ &    $277.2 \pm 8.1$ &   2 &   1 \\[0.5ex] 
    6850029$^{+}$ & $4.44_{-0.04}^{+0.04}$ &  $5640.5_{-99.8}^{+102.5}$ &       $51.1 \pm 4.9$ &        $1.73_{-0.19}^{+0.20}$ & $-0.01_{-0.16}^{+0.14}$ & $0.94_{-0.07}^{+0.06}$ &   $473.5 \pm 12.1$ &   2 &   1 \\[0.5ex] 
    11028172 & $4.21_{-0.04}^{+0.04}$ & $5654.9_{-106.1}^{+110.7}$ &       $49.8 \pm 5.7$ &        $1.70_{-0.22}^{+0.22}$ &  $0.03_{-0.17}^{+0.16}$ & $0.97_{-0.07}^{+0.07}$ &   $387.9 \pm 10.2$ &   2 &   1 \\[0.5ex] 
    4265609 & $4.34_{-0.04}^{+0.04}$ & $5932.0_{-111.4}^{+113.8}$ &       $42.1 \pm 1.4$ &        $1.69_{-0.12}^{+0.12}$ &  $-0.05_{-0.03}^{+0.03}$ & $1.03_{-0.07}^{+0.07}$ &   $466.8 \pm 11.4$ &   1 &   1 \\[0.5ex] 
    7908114 & $4.25_{-0.06}^{+0.05}$ & $5767.1_{-118.8}^{+124.3}$ &       $46.1 \pm 4.7$ &        $1.68_{-0.21}^{+0.21}$ &  $0.03_{-0.17}^{+0.14}$ & $1.00_{-0.08}^{+0.08}$ &   $284.5 \pm 10.9$ &   2 &   1 \\[0.5ex] 
    7428723$^{+}$ & $4.40_{-0.05}^{+0.04}$ & $5826.8_{-105.5}^{+107.0}$ &       $44.2 \pm 5.3$ &        $1.67_{-0.22}^{+0.22}$ & $-0.01_{-0.16}^{+0.14}$ & $1.00_{-0.07}^{+0.07}$ &   $606.2 \pm 15.6$ &   2 &   1 \\[0.5ex] 
    (7189915)$^{+}$ & $4.43_{-0.05}^{+0.04}$ & $5739.0_{-101.5}^{+104.8}$ &       $(44.5 \pm 5.0)$ &        $(1.60_{-0.20}^{+0.20})$ & $-0.02_{-0.15}^{+0.14}$ & $0.97_{-0.07}^{+0.06}$ &   $453.4 \pm 14.0$ &   2 &   1 \\[0.5ex] 
    \hline
    \hline
  \end{tabular}
  \tablefoot{We recapitulate here the main parameters for the anti-solar candidates at solar metallicity. From the left, we show the \textit{Kepler} Input Catalog (KIC) identification, the surface gravity logarithm log$_{10}$g, the effective temperature $T_{\rm eff}$, the rotation period $P_{\rm rot}$, the normalized fluid Rossby number value from Eq.~\ref{eq:RofTeffProt}, the metallicity index ${\rm [Fe/H]}$, the mass $M_*$ , and the photometric magnetic activity proxy $S_{\rm ph}$. The metallic flag Zfl indicates the origin of the quoted [Fe/H] value (APOGEE (0), LAMOST (1), or \citetalias{bergerGaiaKeplerStellarProperties2020} (2)). The last flag Sfl is a sample flag indicating whether the target is in the optimistic (1) or conservative (2) sample. Values  within parentheses are further discussed in the text, and the exponent on the KIC indicates either a solar analog (+) or a seismic target (*).}
\end{table*}

Using Eq.~\ref{eq:RofTeffProt}, we selected all targets with $Ro_{\rm f}/{\rm Ro_{\rm f,\odot}}$ above a given threshold. The anti-solar DR is expected to appear in a regime in which the dominant convective motions that transport angular momentum within the CZ are weakly influenced by the global rotation of the star (see Sect.~\ref{sec:DRprofiles}). To do this, we placed a threshold at $Ro_{\rm f}\geq1.3$ to ensure that we were in this regime. For instance, $Ro_{\rm f}=1.24$ is the lowest $Ro_{\rm f}$ found in \citetalias{Brun+2022} models experiencing an anti-solar regime. Then, we considered two possible values ${\rm Ro_{\rm f,\odot}}=\{0.6;0.9\}$ according to the current uncertainties on solar convection, known as the convective conundrum \citep{hanasogeSolarDynamicsRotation2015}. The lower value 0.6 corresponds to the value needed to obtain a decadal solar-type magnetic cycle in the MHD numerical simulations of \citetalias{Brun+2022}, and 0.9 is the value that is effectively obtained in the solar regime in this same set of simulations. This gave us two thresholds $Ro_{\rm f}/{\rm Ro_{\rm f,\odot}}=\{2.167;1.444\}$, which define two samples that we refer to as the conservative and optimistic sample respectively. To ensure the reliability of this method, we computed the uncertainties and took them into account
. Hence, a target was selected only if its normalized Rossby value from Eq.~\ref{eq:RofTeffProt} had a lower limit above the given threshold.

This selection gave us 62 targets in the optimistic sample, 12 of which belonged to the conservative sample. We individually inspected them using the following method. We first considered the field of view (FoV) in the \emph{Kepler} Asteroseismic Science Operations Center (KASOC) database\footnote{The KASOC database can be accessed at \href{http://kasoc.phys.au.dk}{http://kasoc.phys.au.dk}} \citep{2010AN....331..966K}.
\begin{enumerate}
    \item If the considered target was within the halo of a brighter star, we did not consider it because the $T_{\rm eff}$ or $P_{\rm{rot}}$ measurements might be biased.
    \item We then ensured that no neighboring stars lay in a FoV of $40''$ centered on the target. The image scale of \emph{Kepler} is 3.98 arcseconds per pixel, and the total acquisition FoV is a square of 10 pixels at a side \citep{2016ksci.rept....1V}. If neighbors with a similar magnitude were found, we considered the photometric aperture to ensure that the light curve of the considered target was not polluted. If this was not the case, we did not consider it.
    \item Next, we studied the shape of the target in the FoV to avoid considering close binary systems.
    \item Finally, we visually inspected the light curve and respective rotation diagnostics as indicated in Appendix \ref{sec:FineCheck} in order to ensure the reliability of the detected $P_{\rm rot}$ value.
\end{enumerate}
The different results of these inspections are summarized for each target at solar metallicity in Appendix \ref{sec:FineCheck}.

\begin{figure}
    \centering
    \includegraphics[width=\linewidth]{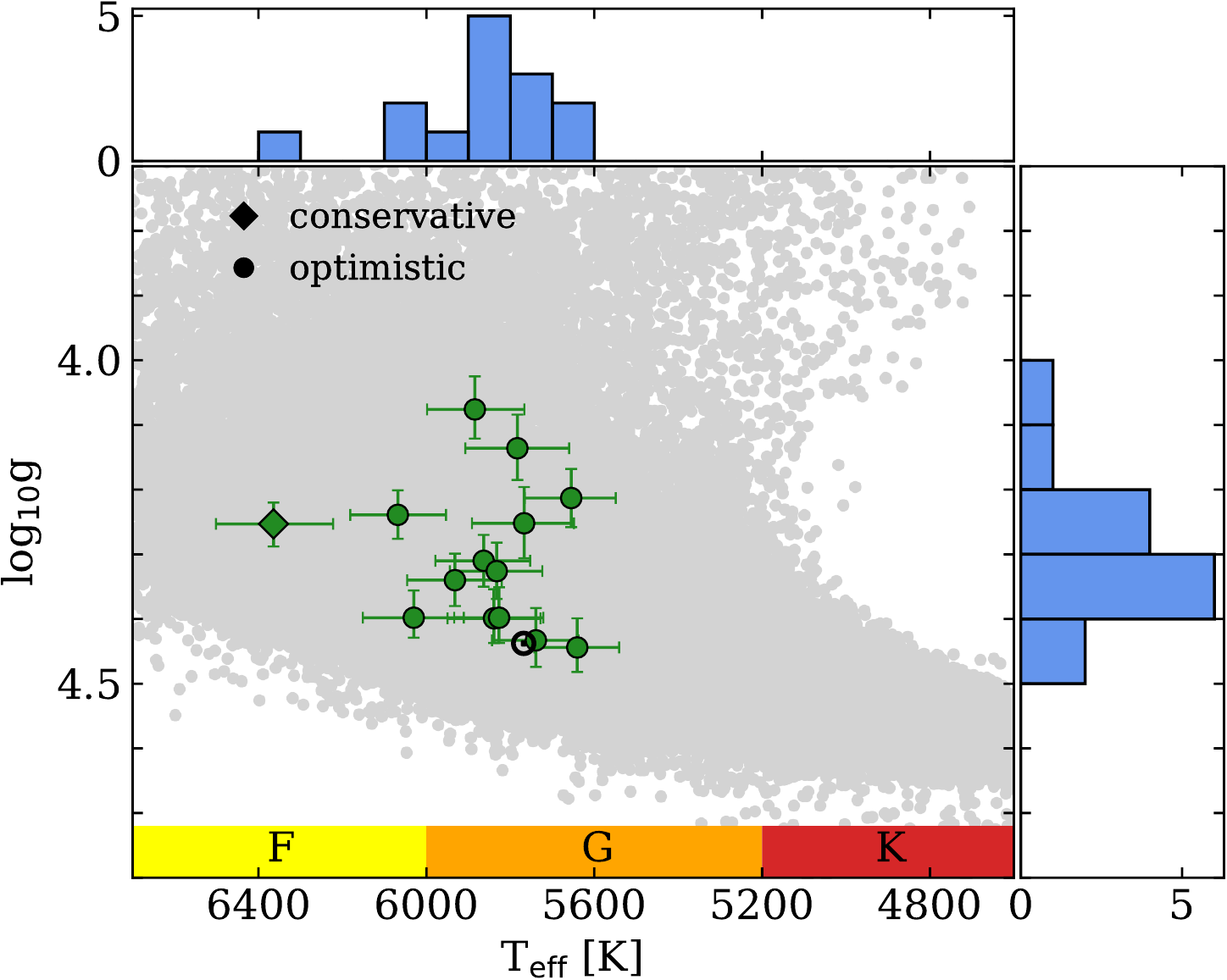}
    \caption{Kiel diagram ($T_{\rm eff}$ vs. log$_{10}$g) for anti-solar DR candidates at solar metallicity. The conservative and optimistic samples are illustrated with green diamonds and circles, respectively. The $\text{circled dot}$  represents the Sun, and spectral types are indicated on the $T_{\rm eff}$-axis. Spectral types are defined here with the standard Harvard Spectral Classification (see, e.g., \citealt{1981A&AS...46..193H} for more details). Histograms on the x- and y-axes correspond to the $T_{\rm eff}$ and log$_{10}$g distributions for the candidates shown in green. We illustrate all rotators from the \citetalias{2019ApJS..244...21S}-\citetalias{santosSurfaceRotationPhotometric2021} catalog with gray dots for comparison.}
    \label{fig:BothSampleZSun}
\end{figure}

After ensuring the reliability of the computed fluid Rossby value with previous inspections, we found 14 stellar candidates at solar metallicity that were likely to be in an anti-solar DR regime. We illustrate them in a Kiel diagram in Fig.~\ref{fig:BothSampleZSun}.
Only one of the 14 candidates that are in the optimistic sample is in the conservative sample (diamond marker). We plot the distributions including the 14 targets on both axes, which show that high $Ro_{\rm f}$ targets are likely to be found around the medians log$_{10}g=4.32$ and $T_{\rm eff}=5836$ K (early-G spectral type) for old MS cool stars with solar metallicity. We summarize the stellar parameters of each target in Table~\ref{tab:ZsunTargets}.

We first note that all anti-solar candidates have rotational periods $P_{\rm rot}$ between $42.1$ and $51.1$ days, effective temperatures $T_{\rm eff}$ from $5640$ to $6365$ K (early-G and late F-type stars), and masses $M_*$ ranging from $0.94$ to $1.22$ $M_\odot$. Because all stellar gravities, log$_{10}$g, lie between $4.08$ and $4.44$~dex, these targets seem consistent with old MS stars. We then note that KIC~6068129, 6850029, and 7428723 can be considered as solar analogs according to their stellar parameters. We considered stars as solar analogs when their parameters (including uncertainties) fit within ranges of $250$ K around ${\rm T_{\rm eff,\odot}}=5772$ K and $0.10$ dex around log$_{10}$g$_\odot=4.44$. Slightly less conservative definitions can be adopted. We refer to \cite{cayreldestrobelStarsResemblingSun1996} for an extensive discussion. Another noteworthy target is KIC~12117868, which is a seismic target \citep{garciaRotationMagnetismKepler2014}. Next, we verified the value of the $S_{\rm ph}$ index, which is a proxy of photometric magnetic activity computed as defined in \citet{mathurPhotometricMagneticactivityMetrics2014}.  $S_{\rm ph}$ of these stars reaches a maximum of $751.6$~ppm and mostly lies in the solar range S$_{\rm ph, \odot}=50-600$~ppm \citep{salabertPhotosphericActivitySun2017}
. The maximum $Ro_{\rm f}/{\rm Ro_{\rm f,\odot}}$ value for these stars is 2.44 and is reached by KIC~10907436, which thus seems to be the most promising candidate for showing anti-solar DR in the sample of stars at solar metallicity.

In Table~\ref{tab:ZsunTargets} we consider KIC~7189915, which should have been excluded from the optimistic sample. This target was first selected thanks to its high $Ro_{\rm f}$ value, which is partly due to a relatively long $P_{\rm rot}=51.1\pm4.7$ days given by \citetalias{2019ApJS..244...21S}-\citetalias{santosSurfaceRotationPhotometric2021}. During individual inspections of it, we revised this value and now consider $P_{\rm rot}=45\pm5$ days, which yields $Ro_{\rm f}/{\rm Ro_{\rm f,\odot}}=1.60^{+0.20}_{-0.20}$ when Eq.~\ref{eq:RofTeffProt} is recomputed with the new period. Consequently, the new $Ro_{\rm f}/{\rm Ro_{\rm f,\odot}}$ lower limit (1.40) is slightly below the threshold $1.3/0.9=1.44$, and we should accordingly strictly not consider this target in Table~\ref{tab:ZsunTargets}. However, we chose to include
it because of its relatively high $Ro_{\rm f}/{\rm Ro_{\rm f,\odot}}$ value and because we are confident through visual inspection of the target that this new period value is reliable. This makes it a promising candidate. Most importantly, however, its parameters are very close to the solar parameters, which makes it an interesting target to investigate because it is probably older than the Sun according to the gyrochronology. As a first guess of its age, we used luminosity, effective temperature, metallicity, and rotation period to interpolate in the grid  of \cite{amardFirstGridsLowmass2019} and found an estimate of about $9.5\pm0.5$ Gyr.

\section{Extended sample at different metallicities}\label{sec:4}
After applying our data-processing steps and $Ro_{\rm f}$ selection to 16,258 solar metallicity targets, we extended our interest to stars with other metallicities and considered all 50,656 rotators in subsample I of \citetalias{2019ApJS..244...21S}-\citetalias{santosSurfaceRotationPhotometric2021}. This implies that the variations in $Ro_{\rm f}$
as a function of metallicity were taken into account. We expressed this dependence as a function of the metallicity index ${\rm [Fe/H]}$ in order to keep the $Ro_{\rm f}$ formula in terms of observable quantities.

\subsection{Fluid Rossby number including metallicity}
In this subsection, we add the metallicity index ${\rm [Fe/H]}$ to the analytical development of the structural fluid Rossby number. Because homology relations (Eq.~\ref{eq:ScalingLawsL&R}) should also depend on the composition \citep{kippenhahnStellarStructureEvolution2012}, we adapted them by explicitly adding ${\rm [Fe/H]}$, such that
\begin{equation}
    L_*\propto M_*^{m_1}({\rm [Fe/H]}+2)^{m_2}\hspace{0.5cm}\text{and}\hspace{0.5cm}R_*\propto M_*^{n_1}({\rm [Fe/H]}+2)^{n_2},
    \label{eq:FeHscaling}
\end{equation}
with $m_2$ and $n_2$ undetermined for now. Moreover, the metallicity impacts the structure of the star. A higher abundance of heavy elements increases the opacity in the star, making radiative transfer less effective. This results in an increase in the temperature gradient \citep{maederPhysicsFormationEvolution2009}. This leads to a deeper convective envelope and a cooler star \citep{karoffInfluenceMetallicityStellar2018,amardFirstGridsLowmass2019} and slower convective speeds at the BCZ, which subsequently impacts the Rossby number \citep{amardImpactMetallicityEvolution2020,seePhotometricVariabilityProxy2021}. We therefore took the dependence on ${\rm [Fe/H]}$ in the structural aspect into account as well by parameterizing $S(\alpha,\beta)$ such that
\begin{equation}
    S(\alpha,\beta)\propto M_*^{q_1}({\rm [Fe/H]}+2)^{q_2}
    \label{eq:SabFeH}
.\end{equation}
Similarly, we obtained
\begin{equation*}
    Ro_{\rm f}\propto M_*^{a_1}({\rm [Fe/H]}+2)^{a_2}\Omega^{-1}
\end{equation*}
\begin{equation}
    \label{eq:RofMZ}
    \text{with}\hspace{0.5cm}\left\{
    \begin{array}{ll}
        a_1=q_1+\dfrac{m_1-(2n_1+1)}{3}\\
        \\
        a_2=q_2+\dfrac{m_2-2n_2}{3}\;
    \end{array}
    \right\}
.\end{equation}
Substituting Eqs.~\ref{eq:FeHscaling} into Eqs.~\ref{eq:L*} and \ref{eq:RofMZ} gives
\begin{equation}
    M_*\propto T_{\rm eff}^{4/(m_1-2n_1)}\left({\rm [Fe/H]}+2\right)^{(2n_2-m_2)/(m_1-2n_1)},
\end{equation}
\begin{equation*}
    Ro_{\rm f}\propto T_{\rm eff}^{b_1}({\rm [Fe/H]}+2)^{b_2}\Omega^{-1}
\end{equation*}
\begin{equation}
    \label{eq:RofTZ}
    \text{with}\hspace{0.5cm}\left\{
    \begin{array}{ll}
        b_1=\dfrac{4a_1}{m_1-2n_1}\\
        \\
        b_2=a_1\dfrac{2n_2-m_2}{m_1-2n_1}+a_2\;
    \end{array}
    \right\}
.\end{equation}
Multiparameter regression fits on the evolutionary tracks of \citetalias{amardFirstGridsLowmass2019} (see fits and values in Fig.~\ref{fig:ScalingLaws}) gave $m_1$, $n_1$ , and $p_1$ equivalent to what was found in Sect.~\ref{sec:3}. In addition, we find $m_2=-1.00\pm0.04$, $n_2=0.01\pm0.01$, and $q_2=-0.81\pm0.08$. It therefore appears that the stellar luminosity $L_*$ is inversely proportional to the metallicity index ${\rm [Fe/H]}$. This is due to the increase in opacity with higher ${\rm [Fe/H]}$ because of the presence of elements with several atomic levels or ionization states. It also appears that the stellar radius $R_*$ is nearly independent of the metallicity, regardless of the mass $M_*$ considered (see Fig.~\ref{fig:ScalingLaws}). We further note that the structure factor $S$ decreases with ${\rm [Fe/H]}$, which means that the CZ thickness increases with metallicity. 
Introducing these values in Eqs.~\ref{eq:RofMZ} and \ref{eq:RofTZ} leads to
\begin{equation*}
    Ro_{\rm f}\propto\dfrac{M_*^{1.93\pm0.05}({\rm [Fe/H]}+2)^{-1.15\pm0.08}}{\Omega_*}\hspace{0.5cm}\text{and}
\end{equation*}
\begin{equation}
    Ro_{\rm f}\propto\dfrac{T_{\rm eff}^{3.29\pm0.09}({\rm [Fe/H]}+2)^{-0.31\pm0.09}}{\Omega_*}.
    \label{eq:RofZAN}
\end{equation}
We find that the $Ro_{\rm f}$ value is inversely proportional to the metallicity index ${\rm [Fe/H]}$, which is consistent with what has been observed by \cite{seePhotometricVariabilityProxy2021}.

We note that $Ro_{\rm f}$ is more sensitive to metallicity when it is expressed as a function of the stellar mass $M_*$ rather than the effective temperature $T_{\rm eff}$. We can understand this through \citetalias{amardFirstGridsLowmass2019}'s evolutionary tracks: when we fix $T_{\rm eff}$ and increase ${\rm [Fe/H]}$ from $-1$ to $+0.3$, there is no significant change in structure or convective power (1\% difference in $r_{\rm BCZ}/R_*$ and a difference of a factor of 2 in the mean convective velocity), hence no strong change in $Ro_{\rm f}$. However, when we now fix $M_*$ and apply the same increase in ${\rm [Fe/H]}$, we note a significant decrease in convective power, leading to a drop in $Ro_{\rm f}$ by almost a factor of 10. Hence, the fluid Rossby number is more sensitive to metallicity effects when it is expressed with $M_*$.

Similarly to Sect.~\ref{sec:RofStruct}, it is convenient to consider a solar reference such that
\begin{equation}
    \dfrac{Ro_{\rm f}}{\rm Ro_{\rm f,\odot}} = \left(\dfrac{P_{\rm rot,*}}{\rm P_{\rm rot,\odot}}\right)\times\left(\dfrac{T_{\rm eff}}{\rm T_{\rm eff,\odot}}\right)^{3.29}\times\left(\dfrac{{\rm [Fe/H]}+2}{2}\right)^{-0.31}.
    \label{eq:RofTeffProtFeH}
\end{equation}
We now applied the same Rossby criterion as in the previous section.

\subsection{Most promising targets at different metallicities}
\begin{table*}
  \centering
  \caption{List and stellar parameters of anti-solar candidates beyond the solar metallicity range}
  \label{tab:ZothTargets}
  \begin{tabular}{lccccrclcc}
    \hline
    \hline\\ [-2.0ex]
    KIC &          log$_{10}$g &          $T_{\rm eff}$ [K] & $P_{\rm rot}$ [days] & $Ro_{\rm f}/{\rm Ro_{\rm f,\odot}}$ &                  [Fe/H] &      $M_*$ [$M_\odot$] & $S_{\rm ph}$ [ppm] & Zfl & Sfl \\[0.5ex]
    \hline
    \hline\\ [-2.0ex]
    2161400 & $4.44_{-0.04}^{+0.04}$ & $6013.9_{-117.3}^{+120.4}$ &       $48.7 \pm 5.6$ &              $2.19_{-0.29}^{+0.29}$ &  $-0.39_{-0.06}^{+0.06}$ & $0.93_{-0.06}^{+0.06}$ &    $253.3 \pm 7.7$ &   1 &   1 \\[0.5ex] 
    6842602 & $4.56_{-0.02}^{+0.01}$ & $6130.5_{-105.0}^{+110.7}$ &       $40.9 \pm 5.2$ &              $2.07_{-0.29}^{+0.30}$ & $-0.66_{-0.15}^{+0.14}$ & $0.90_{-0.03}^{+0.02}$ &   $409.3 \pm 13.8$ &   2 &   1 \\[0.5ex] 
    7097538 & $4.40_{-0.04}^{+0.04}$ & $5875.1_{-109.2}^{+111.7}$ &       $47.7 \pm 6.5$ &              $1.91_{-0.29}^{+0.29}$ &  $-0.20_{-0.04}^{+0.04}$ & $0.94_{-0.07}^{+0.07}$ &    $370.8 \pm 9.7$ &   1 &   1 \\[0.5ex] 
    3628076 & $4.43_{-0.04}^{+0.04}$ & $5879.5_{-120.5}^{+121.6}$ &       $43.3 \pm 5.8$ &              $1.82_{-0.27}^{+0.27}$ &  $-0.43_{-0.02}^{+0.02}$ & $0.87_{-0.06}^{+0.06}$ &   $545.7 \pm 13.6$ &   1 &   1 \\[0.5ex] 
    3113301 & $4.11_{-0.04}^{+0.03}$ & $5853.5_{-101.2}^{+105.7}$ &       $47.7 \pm 6.8$ &              $1.76_{-0.27}^{+0.27}$ &   $0.29_{-0.03}^{+0.03}$ & $1.26_{-0.10}^{+0.05}$ &    $308.3 \pm 7.1$ &   1 &   1 \\[0.5ex] 
    10516429 & $4.50_{-0.04}^{+0.03}$ & $6036.6_{-148.2}^{+131.6}$ &       $(36.6 \pm 4.2)$ &              $(1.75_{-0.25}^{+0.24})$ &  $-0.64_{-0.05}^{+0.05}$ & $0.86_{-0.06}^{+0.05}$ &    $204.2 \pm 5.1$ &   1 &   1 \\[0.5ex] 
    6356292 & $4.47_{-0.04}^{+0.04}$ & $5862.6_{-119.8}^{+126.7}$ &       $38.4 \pm 3.3$ &              $1.68_{-0.19}^{+0.19}$ &  $-0.67_{-0.05}^{+0.05}$ & $0.82_{-0.05}^{+0.06}$ &   $354.3 \pm 13.6$ &   1 &   1 \\[0.5ex] 
    6436380 & $4.22_{-0.04}^{+0.03}$ & $5999.2_{-116.0}^{+115.3}$ &       $(40.0 \pm 2.6)$ &              $(1.62_{-0.15}^{+0.15})$ &   $0.16_{-0.02}^{+0.02}$ & $1.19_{-0.08}^{+0.06}$ &    $348.1 \pm 7.7$ &   1 &   1 \\[0.5ex] 
    \hline
    \hline
  \end{tabular}
  \tablefoot{ Similar to the Table~\ref{tab:ZsunTargets} for promising targets beyond the solar metallicity range. The normalized fluid Rossby number value is computed from Eq.~\ref{eq:RofTeffProtFeH}.}
\end{table*}
Following the method we described in Sect.~\ref{sec:PMS-SG}, we applied similar steps of data processing to the whole \citetalias{2019ApJS..244...21S}-\citetalias{santosSurfaceRotationPhotometric2021} subsample I. First, we obtained a maximum value $Ro_{\rm f}/{\rm Ro_{\rm f,\odot}}$=0.608 when applying Eq.~\ref{eq:RofTeffProtFeH} to PMS phases to the evolutionary tracks of \citetalias{amardFirstGridsLowmass2019}. This again confirms that no explicit filtering of PMS stars is necessary. Then, the filtering we applied to exclude subgiants was adapted for each metallicity available in \citetalias{amardFirstGridsLowmass2019} (see details in Appendix \ref{sec:TAMSvsZ}), which allowed us now to consider the range of metallicities $-1 < {\rm [Fe/H]} < +0.3$\,dex. After this data processing, we obtained 37,237 targets on which we applied Eq.~\ref{eq:RofTeffProtFeH}. We clearly found again all targets of Table~\ref{tab:ZsunTargets} in the optimistic sample\footnote{Applying Eq.~\ref{eq:RofTeffProtFeH} on targets in Table~\ref{tab:ZsunTargets} does not significantly change their $Ro_{\rm f}/Ro_{\rm f,\odot}$ values, as their [Fe/H] amplitude is small. The strongest change is experienced by KIC~10907436, which fluid Rossby increases from 2.44 to 2.47.}, and extended it by adding 37 new 
potential candidates. One of them was also in the conservative sample.
\begin{figure}
    \centering
    \includegraphics[width=\linewidth]{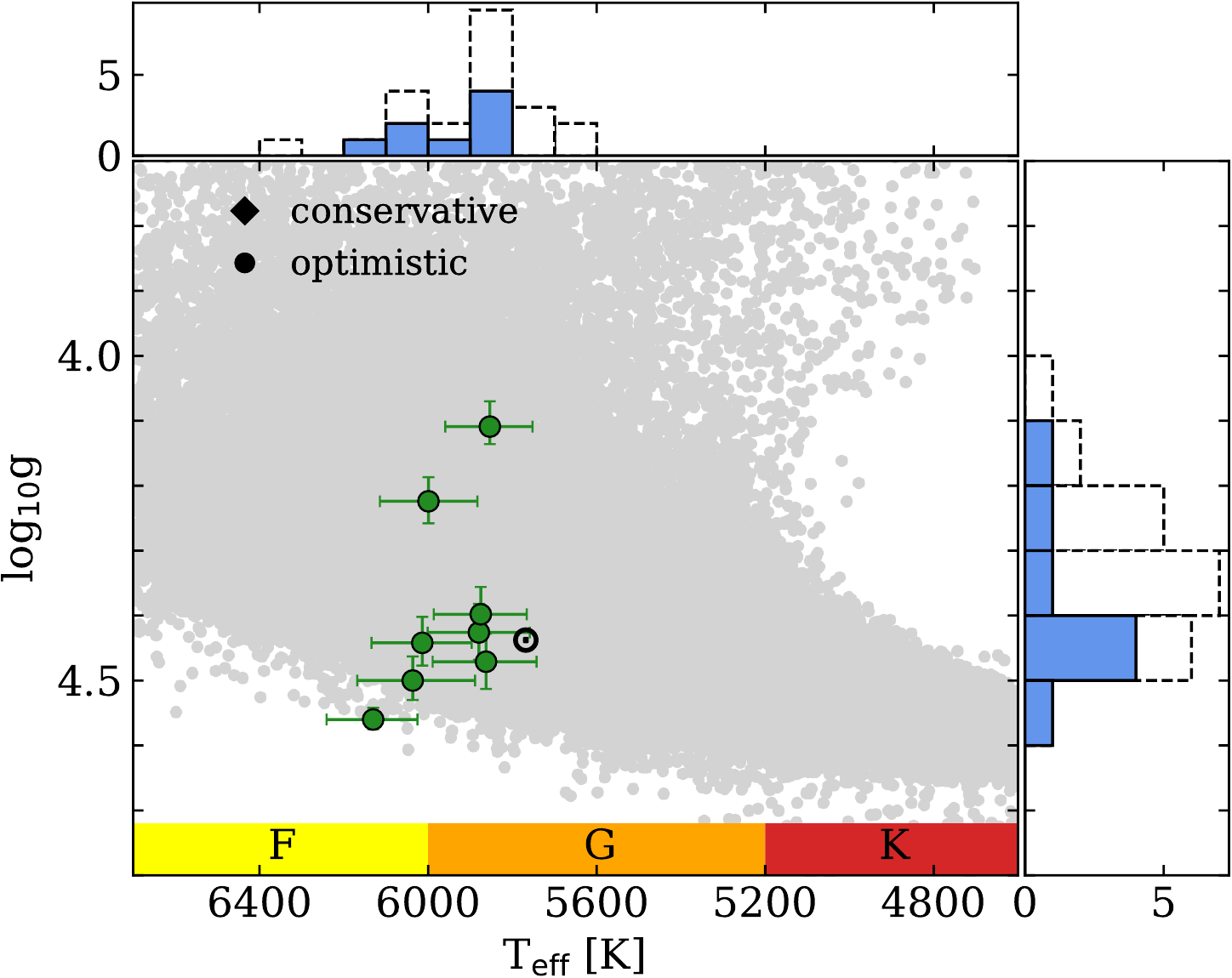}
    \caption{Similar to Fig.~\ref{fig:BothSampleZSun} for anti-solar DR candidates beyond the solar metallicity range, summarized in Table~\ref{tab:ZothTargets}. All of them are part of the optimistic sample. They are illustrated with green circles. Blue histograms on the x- and y-axes are proposed for the green dot sample. For comparison, we add dotted black histograms, which represent the whole set of candidates for all metallicities (i.e., the targets in Table~\ref{tab:ZsunTargets} and \ref{tab:ZothTargets} ).}
    \label{fig:BothSampleZOth}
\end{figure}
Again we performed the individual visual inspections described in Sect.~\ref{sec:ZsunCandidates} and summarize the details in Appendix \ref{sec:FineCheck}. These inspections lead to eight new candidates, which are listed in Table~\ref{tab:ZothTargets} and illustrated in Fig.~\ref{fig:BothSampleZOth}.

Table~\ref{tab:ZothTargets} shows that these eight new candidates are only in the optimistic sample. They show a rotation period between 36.6 and 48.7 days, $T_{\rm eff}$ from 5853 to 6131\,K, $M_*$ between 0.82 and 1.26\,$M_\odot$, and [Fe/H] ranging from -0.67 to +0.29\,dex. The $S_{\rm ph}$ values for all candidates fall within the solar values of 50 and 600 ppm at the minimum and maximum activity, respectively \citep{salabertPhotosphericActivitySun2017}. We note that KIC~3113301 and 6436380 show low log$_{10}$g while still being on the MS as they have a higher metallicity for a given $T_{\rm eff}$. In contrast, the low-metallicity star KIC~6842602 shows a significantly high log$_{10}$g. Finally, we also considered a new $P_{\rm rot}$ for KIC~10516429 and 6436380 compared to the original catalog \citetalias{2019ApJS..244...21S}-\citetalias{santosSurfaceRotationPhotometric2021}, resulting from a new period determination that was performed during the individual visual inspections. Original rotation periods in \citetalias{2019ApJS..244...21S}-\citetalias{santosSurfaceRotationPhotometric2021} are $47.7 \pm 6.7$ and $50.4 \pm 9.0$ days. Applying Eq.~\ref{eq:RofTeffProtFeH} to these targets with these new rotation periods, we still obtain $Ro_{\rm f}$ that is high enough to consider them in the optimistic sample.

Fig.~\ref{fig:BothSampleZOth} shows the eight candidates on a Kiel diagram. Blue histograms illustrate their distribution, located around the medians $log_{10}g=4.43$ and $T_{\rm eff}=5939$ K. We also show the distribution of the whole optimistic (for all metallicities) sample with dotted black histograms. This distribution is located around the medians $log_{10}g=4.37$ and $T_{\rm eff}=5869$ K, which is thus the parameter range in which $Ro_{\rm f}$ is highest and in which the anti-solar DR regime is likely to be reached for MS cool-stars for different metallicities.
\begin{figure*}
    \centering
    \includegraphics[width=0.7\linewidth]{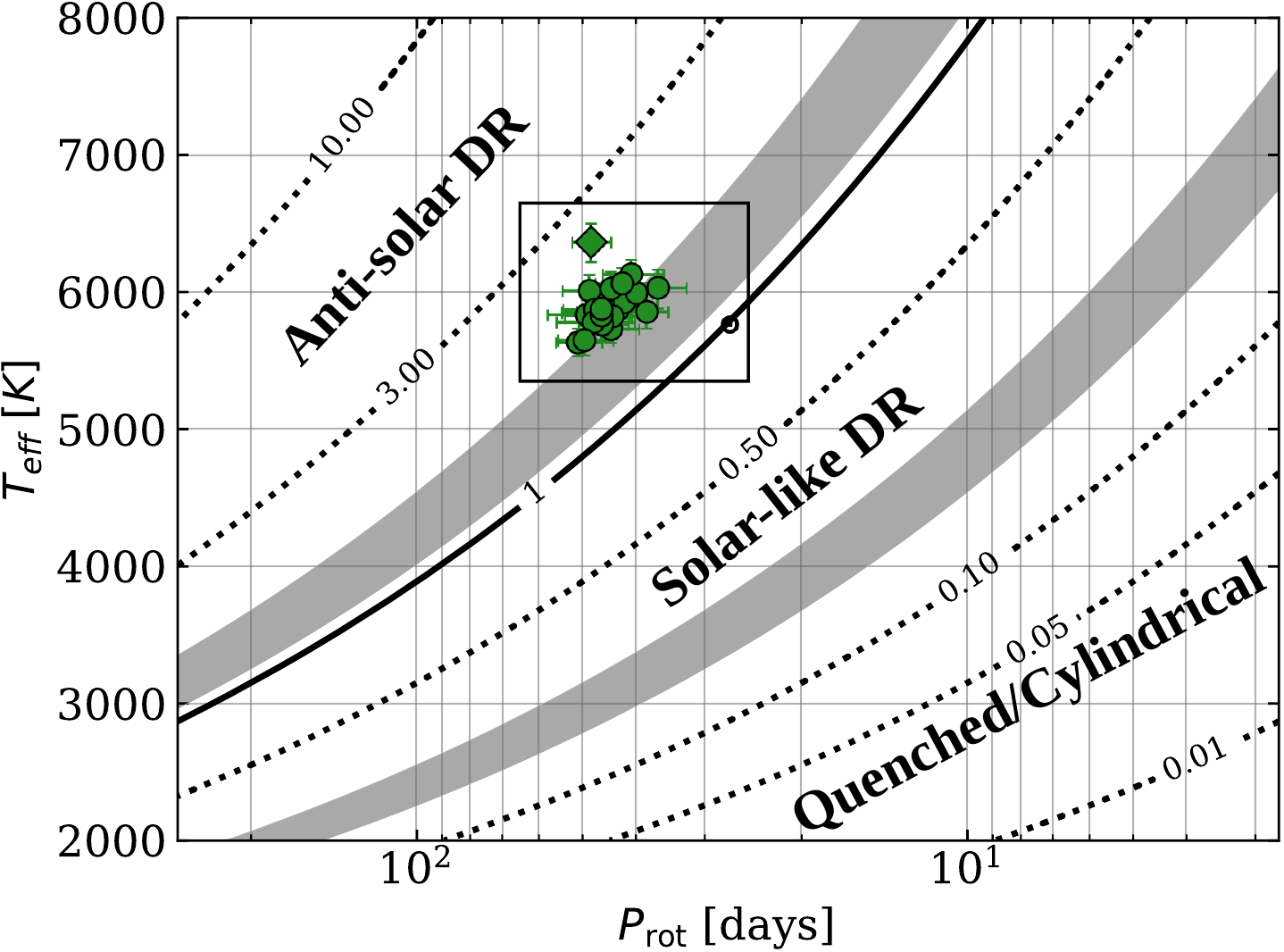}\\
    \vspace{0.5cm}
    \includegraphics[width=0.7\linewidth]{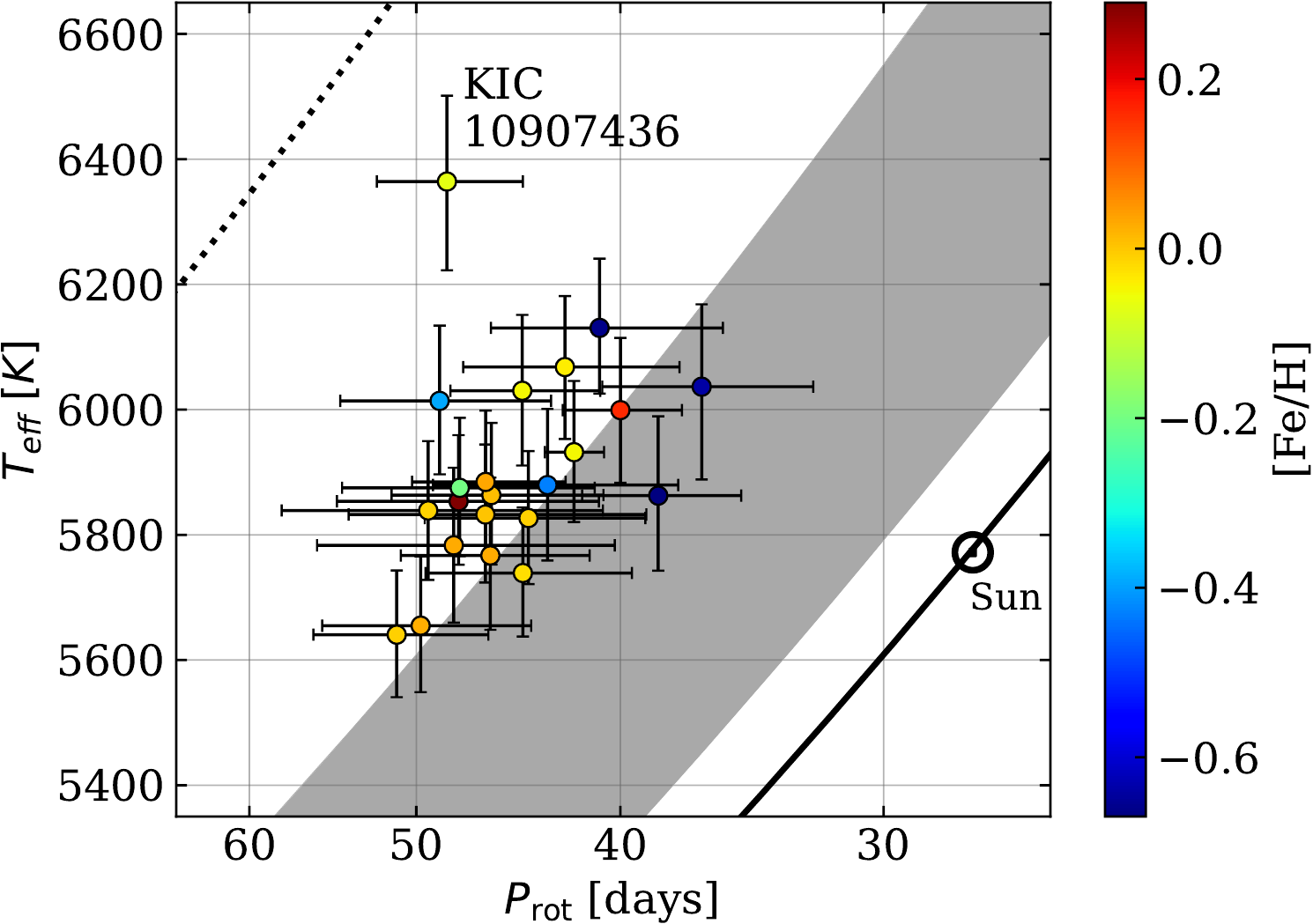}
    \caption{DR diagram in which different DR regimes are expected: anti-solar, solar-type, and quenched or cylindrical (see Fig.~\ref{fig:DRsummary}). Gray regions represent transition regions between DR regimes. \textit{Top:} Dotted lines represent constant $Ro_{\rm f}/{\rm Ro_{\rm f,\odot}}$ as a function of $T_{\rm eff}$ and $P_{\rm rot}$ as deduced from Eq.~\ref{eq:RofTeffProt}. The solid black line represents the value $Ro_{\rm f}/{\rm Ro_{\rm f,\odot}}=1$. \textit{Bottom:} Zoom into the top panel around the anti-solar DR candidates, color-coded by metallicity.}
    \label{fig:BothSampleAllZ}
\end{figure*}

Finally, we illustrate the whole set of the 22 considered optimistic targets in the left panel of Fig.~\ref{fig:BothSampleAllZ} for all metallicities. This is a diagram of $Ro_{\rm f}$ as a function of $T_{\rm eff}$ and $P_{\rm rot}$. The DR regimes presented in Fig.~\ref{fig:DRsummary} are expected in different regions. Two gray transition regions represent the anti-solar or solar-type DR transition (close to $Ro_{\rm f}=1$), and the solar-type or cylindrical transition (close to $Ro_{\rm f}$=0.15, \citealt{brunDifferentialRotationOvershooting2017}, \citeyear{Brun+2022}). The width of the transition is defined between the two solar values ${\rm Ro_{\rm f,\odot}}$ from Sect.~\ref{sec:ZsunCandidates}. This translates into $Ro_{\rm f}/{\rm Ro_{\rm f,\odot}}=(1.67,1.11)$ and $(0.25,0.16)$ . We also show a zoom of this plot in the right panel and indicate the [Fe/H] value for each target on a color bar. We note that most of solar metallicity stars tend to cluster in the lower left corner. We find 4 targets in the transition region toward the anti-solar DR regime, and the other targets lie in the region in which an anti-solar DR regime is expected.

\section{Discussion}\label{sec:5}

We succeeded to obtain 22 stellar candidates that are likely to be in an anti-solar DR regime. We are confident about the robustness of the final sample thanks to different filtering methods and several individual visual inspections for each of the targets presented (see Sect.~\ref{sec:3}). However, some good candidates may have been missed during the different selection processes. They are discussed in this section.

To estimate $Ro_{\rm f}$, we used $P_{\rm rot}$ values taken from the most recent and complete rotational catalog in date for the \textit{Kepler} field (\citetalias{2019ApJS..244...21S}-\citetalias{santosSurfaceRotationPhotometric2021}), where their $P_{\rm rot}$ measurements rely on rotational modulations from starspots in light curves. However, the exact nature of magnetic activity under anti-solar DR is still under debate from a numerical point of view \citep{karakMagneticallyControlledStellar2015,vivianiStellarDynamosTransition2019}, first concerning the possibility of long-term cyclic solutions (\citealt{karakStellarDynamosSolar2020}, \citealt{norazImpactAntisolarDifferential2022} and references therein). Moreover, global numerical simulations currently do not self-consistently resolve starspots. Hence, even if most dynamo simulations highlight stationary configurations, we do not have clues about the starspot formation rate and behavior in a high Rossby number regime. The detection of their modulation in photometric light curves is thus questionable and could be hard, which might explain the relatively small number of promising targets our method selected in the \textit{Kepler} field.

Observations of \cite{brandenburgEnhancedStellarActivity2018} have reported high Rossby targets that show an increase in chromospheric variability compared to the decreasing trend that is expected with the Rossby number. They argued that this enhanced activity for slow rotators could be caused by an anti-solar DR regime. Nevertheless, we would add here that it currently appears to be hard to distinguish whether these strong variabilities are due to an anti-solar DR regime or are caused by tidally triggered magnetic activity in multiple-stars system \citep[e.g.,][]{1991A&A...251..183S,kovariConfinedDynamoMagnetic2021}. These targets are indeed statistically isolated in the parameter space for slow rotators (see the large $S_{\rm ph}$ in Fig.~14 of \citealt{santosSurfaceRotationPhotometric2021}), which could also be an indication for unusually high magnetic activity.\\
We note that the 22 high Rossby targets that were finally retained as promising candidates for anti-solar rotation in our study do not exhibit $S_{\rm ph}$ values that are significantly higher ($<752$~ppm) than those observed on the Sun (see \citealt{salabertPhotosphericActivitySun2017}). We also note that 25 out of 99 targets have higher $S_{\rm ph}$ values ($>1000$ ppm), but they were discarded each time during visual inspections because they were likely to be polluted by a neighbor or by instrumental effects (details are available in Appendix \ref{sec:FineCheck}).

Another explanation for the relatively small number of candidates we found would be a less efficient angular momentum loss for old solar-type stars. A stalled wind-braking mechanism at later stages may prevent stars from entering the high Rossby number regime. This has indeed been observed through photometry \citep{vansadersWeakenedMagneticBraking2016} and asteroseismology techniques \citep{hallWeakenedMagneticBraking2021}. However, this question is still debated in the community, and we refer to Sect. \ref{sec:1} of this paper for other details.

Moreover, the method we proposed to determine the transition between the MS and subgiant phase is a first approximation. Even if linear fits in the parameter space (log$_{10}$g; $T_{\rm eff}$) are good approximations for most of the models taken from \citetalias{amardFirstGridsLowmass2019} (see details in Appendix \ref{sec:TAMSvsZ}), a more accurate characterization of the evolutionary stage of the stars in this transition region would require an asteroseismic analysis. For the same reason, MS cool stars near the end of the MS might have been filtered with our method because of relatively high log$_{10}$g or relatively low $T_{\rm eff}$. Older stars are more likely to be slow rotators, which makes them good candidates for hosting a high Rossby number. Hence, we may have filtered out some promising anti-solar candidates.

Our selection method from the \citetalias{2019ApJS..244...21S}-\citetalias{santosSurfaceRotationPhotometric2021} rotation catalog did not select the anti-solar candidates proposed by \cite{reinholdDiscriminatingSolarAntisolar2015} or \cite{benomarAsteroseismicDetectionLatitudinal2018} as promising targets for hosting anti-solar DR. However, we note that because of the mode suppression by magnetic features, seismic targets tend to be only weakly active \citep[e.g.,][]{Chaplin2011,mathurRevisitingImpactStellar2019}. This also means that detecting the signature of surface rotation due to starspots for these targets might be difficult. Consequently, \citetalias{2019ApJS..244...21S}-\citetalias{santosSurfaceRotationPhotometric2021} were not able to determine photometric surface $P_\text{rot}$ for all seismic targets in \cite{benomarAsteroseismicDetectionLatitudinal2018}. Nevertheless, when we consider the rotation rates that were determined through asteroseismology by the authors, KIC~8938364 and 3427720 become interesting, and would fall in the anti-solar DR region and in the transition (gray area) of Fig.~\ref{fig:BothSampleAllZ}, respectively. Unfortunately, the associated error bars derived from this technique are large and prevent us from considering these targets as promising. We also searched for candidates among the LEGACY/\emph{Kepler} seismic targets of \cite{lundStandingShouldersDwarfs2017} and \cite{hallWeakenedMagneticBraking2021} for which no photometric period was detected and failed to find any targets that clearly passed our selection process.

To be consistent with the mass estimates we considered from \citetalias{2019ApJS..244...21S}-\citetalias{santosSurfaceRotationPhotometric2021}, we took the [Fe/H] values from \citetalias{bergerGaiaKeplerStellarProperties2020}, which come mostly from isochrone fittings, and thus give relatively large error bars. Nevertheless, for half of the final candidates, the [Fe/H] values come from spectroscopic observations, which are known to be more precise in general. We also note that seven of the eight candidates outside the solar metallicity range, for which the $Ro_{\rm f}$ computation takes metallicity into account, have spectroscopic [Fe/H] values (see Table~\ref{tab:ZothTargets}). Therefore, our selection method may have missed targets with inaccurate [Fe/H], but our final sample is likely quite robust in this respect.

\section{Conclusion}\label{sec:6}

In conclusion, we presented a novel approach for highlighting the most promising candidates for the detection of anti-solar DR regime in MS cool stars. We proposed a first application of our method to the \textit{Kepler} field.

We first derived an analytical formula to quantify the fluid Rossby number $Ro_{\rm f}$ from observational quantities ($P_{\rm rot}$, $T_{\rm eff}$) while taking
the structural aspect of solar-type stars into account. After calibrating it with evolutionary tracks of \citetalias{amardFirstGridsLowmass2019}, we then applied this formula to the new \citetalias{2019ApJS..244...21S}-\citetalias{santosSurfaceRotationPhotometric2021} rotation catalog of the \textit{Kepler} field. A first selection was made on stars in the solar metallicity range ($-0.1$\,dex$<{\rm [Fe/H]}<+0.1$\,dex) with a criterion inspired from recent global MHD numerical simulations \citepalias{Brun+2022}. We took care to properly filter pre-MS and post-MS stars in our analysis and finally inspected each selected target to ensure the reliability of the $Ro_{\rm f}$ values we computed.

Next, we extended the derivation of $Ro_{\rm f}$ to add a dependence on the metallicity, which can notably change the structure of stars for a given mass. We adapted our filtering method and considered targets with a metallicity between [Fe/H]=-1 and +0.3\,dex.

This study finally gives us 22 anti-solar DR stellar candidates, 14 of which are at the solar metallicity. We especially note KIC~10907436, which has a significantly high $Ro_{\rm f}$ and is therefore one of the most promising candidates for the anti-solar DR regime. Four solar analogs (KIC~6068129, 6850029, 742872, and 7189915) are also particularly interesting due to their long $P_{\rm rot}$ (between 44.2 and 51.1 days) for the study of the future evolution of the Sun. More generally, all 22 targets possess $P_{\rm rot}$ between 36.6 and 51.1 days for spectral types between $T_{\rm eff}=5641$ and $6364$ K. The determination of their age could therefore give new insights into the debate about the gyrochronology break for old solar-type stars, and more generally, for their rotational evolution \citep{curtisWhenStalledStars2020,davidFurtherEvidenceModified2022}. Another noteworthy candidate is KIC~12117868, which is a seismic target. It could therefore be of interest for future measurements or characterizations, in particular, for its latitudinal rotation profile through asteroseismic inversions \citep{benomarAsteroseismicDetectionLatitudinal2018}, in order to confirm or refute that it has an anti-solar rotation profile.

Anti-solar DR detections on MS cool stars would provide useful observational constraints to better understand the link between rotation and magnetism of slowly rotating stars. Long-term observations would bring constraints on their magnetic activity and would therefore help to determine whether anti-solar differentially rotating stars could sustain magnetic cycles \citep{karakStellarDynamosSolar2020,norazHOWWILLSOLAR2021}. Such constraints for solar analogs would also help to decipher what types of dynamo mechanisms are acting within our Sun \citep{Brun+2022}.

To conclude, this paper was a first attempt to compute $Ro_{\rm f}$ from observations and to apply this method to the \textit{Kepler} field. The analytical development of the fluid Rossby number is generic and could be applied to the field of future missions such as PLAnetary Transits and Oscillations of stars (PLATO, \citealt{rauerPLATOMission2014}).

\begin{acknowledgements}
QN is thankful to J. Ahuir, A. Finley, L. Gizon, S. Jeffers and T. Reinhold for useful discussions. We acknowledge financial support by ERC Whole Sun Synergy grant \#810218., funding by INSU/PNST grant, CNES/Solar Orbiter, CNES/PLATO, and CNES/GOLF funds. ARGS acknowledges the supported by FCT through national funds and by FEDER through COMPETE2020 by these grants: UIDP/04434/2020; PTDC/FIS-AST/30389/2017 \& POCI-01-0145-FEDER-030389. ARGS is supported by FCT through the work contract No. 2020.02480.CEECIND/CP1631/CT0001. SM acknowledges support by the Spanish Ministry of Science and Innovation with the Ramon y Cajal fellowship number RYC-2015-17697 and the grant number PID2019-107187GB-I00.
This paper includes data collected by the \emph{Kepler} mission and obtained from the MAST data archive at the Space Telescope Science Institute (STScI). Funding for the \emph{Kepler} mission is provided by the NASA Science Mission Directorate. STScI is operated by the Association of Universities for Research in Astronomy, Inc., under NASA contract NAS 5–26555. This publication also makes use of data products from the Two Micron All Sky Survey, which is a joint project of the University of Massachusetts and the Infrared Processing and Analysis Center/California Institute of Technology, funded by the National Aeronautics and Space Administration and the National Science Foundation. We also thank the \emph{Kepler} Asteroseismic Science Operations Center, KASOC, and the Aarhus University for maintaining this website. Finally, we thank the anonymous referee for constructive comments that have led to the improvement of the manuscript.
\end{acknowledgements}

\bibliographystyle{biblio/aa}
\bibliography{biblio/RossbyTeff}

\begin{appendix}

\section{TAMS as a function of the metallicity}\label{sec:TAMSvsZ}

\begin{figure*}
    \centering
    \includegraphics[width=0.49\linewidth]{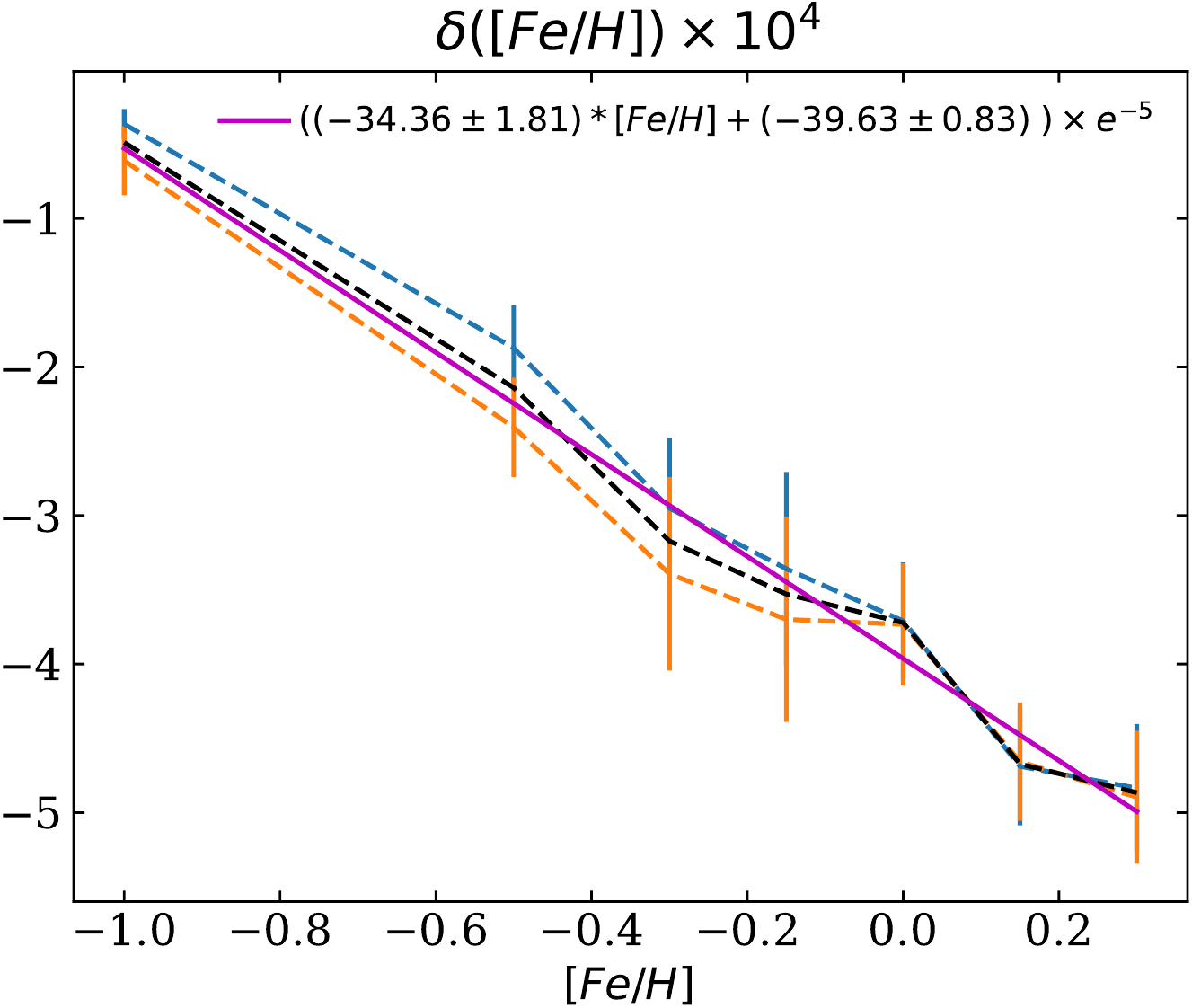}
    \includegraphics[width=0.49\linewidth]{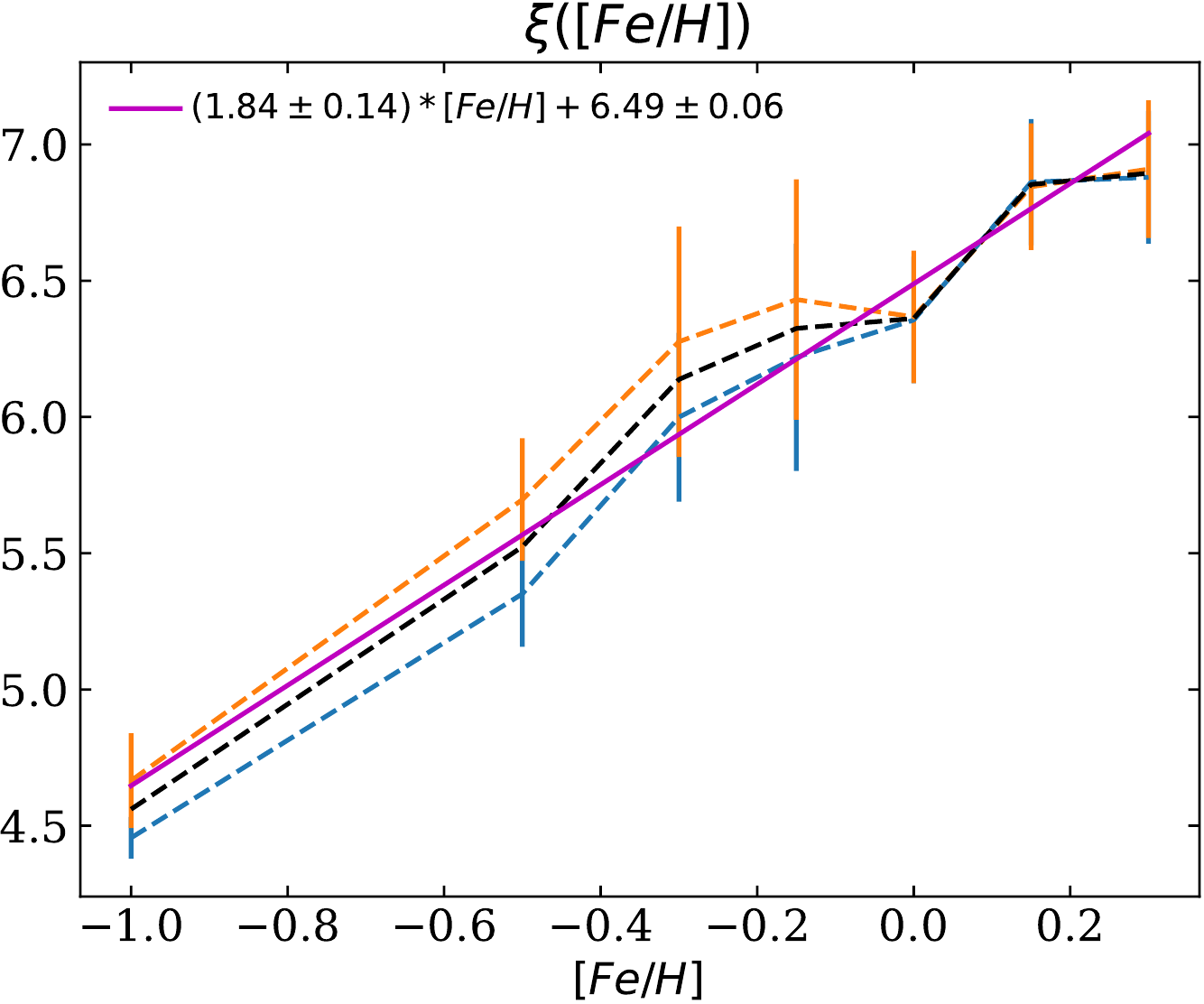}
    \caption{\textit{Left:} $\delta$ slopes from Table~\ref{tab:SBcuts} as a function of ${\rm [Fe/H]}$. Orange and blue correspond to medium and slow rotators, respectively. The dotted black line represents the $\delta$ mean for each metallicity index of the table, and the purple line is a linear regression of these mean values. \textit{Right:} Similar to the left panel for y-intercepts $\xi$.}
    \label{fig:DeltXiFeH}
\end{figure*}

It is a difficult task to observationally distinguish between evolved MS stars and young subgiants. In this situation, an arbitrary choice has to be made in order to filter out stars from the MS populations that might be subgiants. Because we wished to select only cool MS stars in this study, we chose an approach similar to what was considered by \citetalias{2019ApJS..244...21S}-\citetalias{santosSurfaceRotationPhotometric2021}. We thus fit stellar evolution models of \citetalias{amardFirstGridsLowmass2019} at the TAMS as a function of log$_{10}$g and $T_{\rm eff}$ for a given $P_{\rm rot,ini}$ and a given ${\rm [Fe/H]}$. This gives us a limit in a Kiel diagram (e.g., Fig.~\ref{fig:BothSampleZSun}) above which the stars are considered subgiants. We did this for 7 ${\rm [Fe/H]}$ values between $+0.3$ and $-1.0$ for slow and medium rotators. We did not consider the fast rotators because their rotational history was specifically treated for the most massive stars (see Sect.~2.6.1 of \citetalias{amardFirstGridsLowmass2019}). Moreover, we expect anti-solar DR profiles to appear for slowly rotating stars. Slopes $\delta$ and y-intercepts $\xi$ of the TAMS(log$_{10}$g, $T_{\rm eff}$) regressions are reported in Table~\ref{tab:SBcuts}.

\begin{figure}
    \centering
    \includegraphics[width=\linewidth]{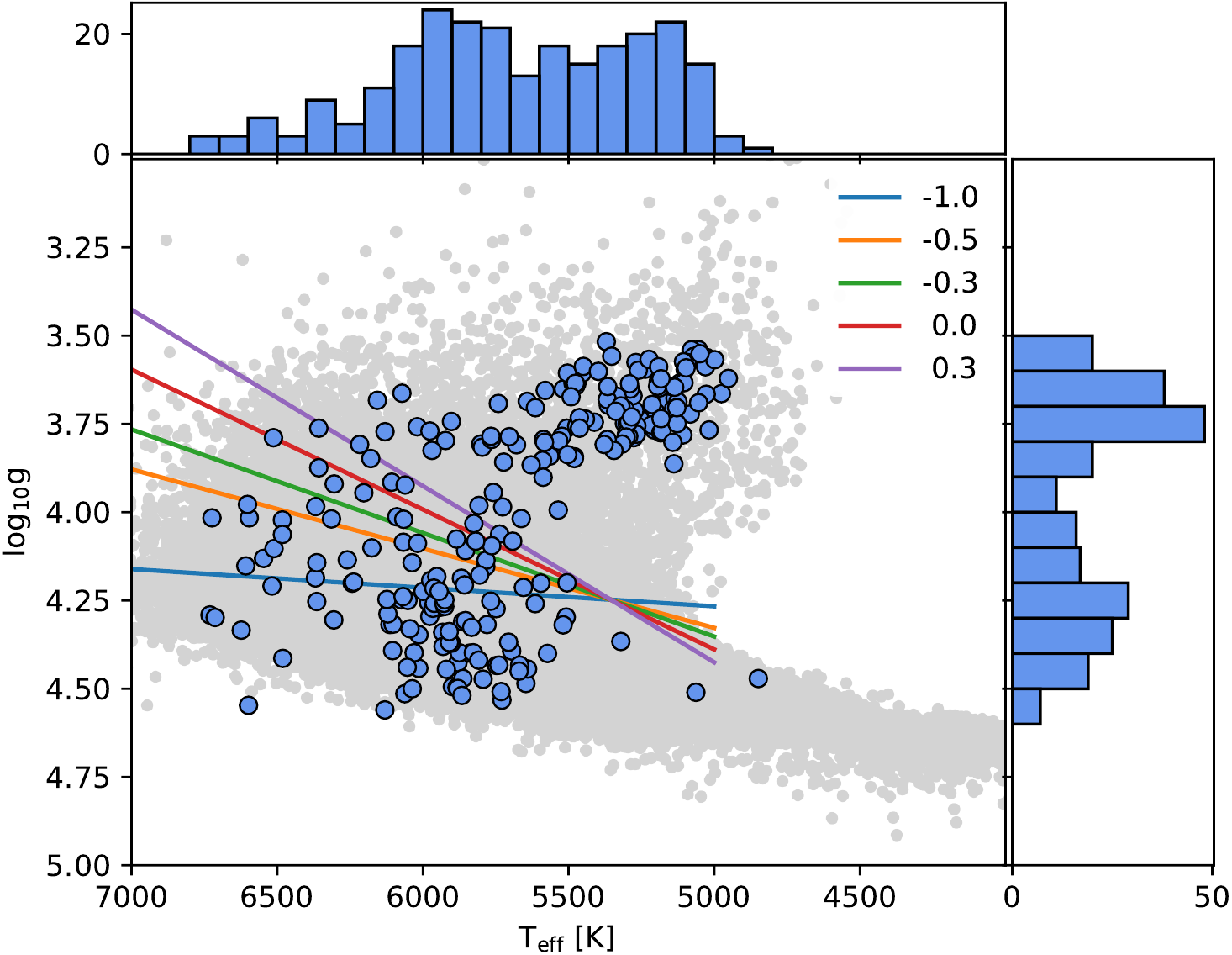}
    \caption{Kiel diagram of the optimistic sample by applying Eq.~\ref{eq:RofTeffProtFeH} to the whole \citetalias{2019ApJS..244...21S}-\citetalias{santosSurfaceRotationPhotometric2021} rotation catalog before filtering masses, metallicities, and subgiants and before individual inspections. Histograms are illustrated in blue on both axes for the selected targets. We show subgiant cuts for different metallicities computed with Eq.~\ref{eq:ABfctZ} in the same diagram.}
    \label{fig:TargetsAndSubgiants}
\end{figure}

We then determined the behavior of $\delta$ and $\xi$ as functions of the metallicity index ${\rm [Fe/H]}$ through linear regressions (see Fig.~\ref{fig:DeltXiFeH}), which gives
\begin{eqnarray}
    \delta({\rm [Fe/H]})&=&((-34.36\pm1.81)\times{\rm [Fe/H]}\nonumber\\
    &\qquad& \qquad + \qquad (-39.63\pm0.83)).10^{-5}\\
    \xi({\rm [Fe/H]})&=&(1.84\pm0.14)\times{\rm [Fe/H]}\nonumber\\
    &\qquad& \qquad + \qquad 6.49\pm0.06 - 0.12 \; .
    \label{eq:ABfctZ}
\end{eqnarray}
We note that the resulting $\xi$ has to be shifted toward low log$_{10}$g values (-0.12 dex) in order to be compatible with log$_{10}$g of \citetalias{2019ApJS..244...21S}-\citetalias{santosSurfaceRotationPhotometric2021}. These two equations are valid for $+0.3\geq{\rm [Fe/H]}\geq-1.0$, that is, the range of metallicities in the \citetalias{amardFirstGridsLowmass2019} evolutionary grid. Finally, we confirm that subgiants do not exist for $T_{\rm eff}<5000K$ and log$_{10}g<4.4$ based on the tracks in this grid of \citetalias{amardFirstGridsLowmass2019}. The Universe is currently too young to consider that MS stars in this temperature range have already left the MS. We finally illustrate some subgiant cuts in Fig.~\ref{fig:TargetsAndSubgiants}.

\begin{table}
\centering
\caption{Parameters of the different linear regression at the TAMS in order to determine subgiant cuts in a Kiel diagram.}
\label{tab:SBcuts}
\begin{tabular}{crcc}
\hline
\hline\\ [-2.0ex]
$P_{\rm rot,ini}$ & [Fe/H] & $\delta$ & $\xi$ \\[0.5ex]
(days) &&& \\
\hline
\hline\\ [-2.0ex]
              9 &  -1.00 & $(-3.64 \pm 1.01)\times 10^{-5}$ & $4.46 \pm 0.08$ \\[0.5ex]
              9 &  -0.50 & $(-1.87 \pm 0.29)\times 10^{-4}$ & $5.35 \pm 0.19$ \\[0.5ex]
              9 &  -0.30 & $(-2.95 \pm 0.47)\times 10^{-4}$ & $6.00 \pm 0.31$ \\[0.5ex]
              9 &  -0.15 & $(-3.36 \pm 0.65)\times 10^{-4}$ & $6.22 \pm 0.42$ \\[0.5ex]
              9 &   0.00 & $(-3.71 \pm 0.39)\times 10^{-4}$ & $6.36 \pm 0.23$ \\[0.5ex]
              9 &   0.15 & $(-4.69 \pm 0.40)\times 10^{-4}$ & $6.86 \pm 0.23$ \\[0.5ex]
              9 &   0.30 & $(-4.84 \pm 0.43)\times 10^{-4}$ & $6.88 \pm 0.24$ \\[0.5ex]
\hline\\ [-2.0ex]
              4.5 &  -1.00 & $(-6.10 \pm 2.33)\times 10^{-5}$ & $4.67 \pm 0.17$ \\[0.5ex]
              4.5 &  -0.50 & $(-2.41 \pm 0.33)\times 10^{-4}$ & $5.70 \pm 0.22$ \\[0.5ex]
              4.5 &  -0.30 & $(-3.39 \pm 0.65)\times 10^{-4}$ & $6.28 \pm 0.42$ \\[0.5ex]
              4.5 &  -0.15 & $(-3.70 \pm 0.69)\times 10^{-4}$ & $6.43 \pm 0.44$ \\[0.5ex]
              4.5 &   0.00 & $(-3.73 \pm 0.41)\times 10^{-4}$ & $6.37 \pm 0.24$ \\[0.5ex]
              4.5 &   0.15 & $(-4.66 \pm 0.40)\times 10^{-4}$ & $6.85 \pm 0.23$ \\[0.5ex]
              4.5 &   0.30 & $(-4.90 \pm 0.45)\times 10^{-4}$ & $6.91 \pm 0.25$ \\[0.5ex]
\hline
\hline
\end{tabular}
\tablefoot{We regrouped the slopes $\delta$ and y-intercepts $\xi$ of the different linear regressions, with the corresponding $P_{\rm rot,ini}$ in days and metallicity index ${\rm [Fe/H]}$.}
\end{table}

\FloatBarrier

\section{Detailed method of individual light-curve inspections and additional rotation diagnostics}\label{sec:FineCheck}
 We detail here the different verifications we performed in the step 4 of individual inspections as discussed in Sect.~\ref{sec:PMS-SG}. For each target in the optimistic sample, we inspected the three KEPSEISMIC light curves used in \citetalias{2019ApJS..244...21S}-\citetalias{santosSurfaceRotationPhotometric2021} and the Presearch Data Conditioning - Maximum A Posteriori \citep[PDC-MAP; e.g.,][]{Jenkins2010,Smith2012,Stumpe2012} light curve. KEPSEISMIC light curves were obtained with {\it Kepler} Asteroseimic Data Analysis and Calibration Software \citep[KADACS;][]{Garcia2011}, which employs customized apertures and corrects for outliers, jumps, drifts, and discontinuities at the edges of the {\it Kepler} quarters. Gaps smaller than 20 days were filled using inpainting techniques \citep{Garcia2014a,Pires2015}. Finally, we applied three filters with cutoff periods at 20, 55, and 80 days, resulting in three KEPSEISMIC light curves per star. It was crucial to filter the data to remove instrumental modulations, but it can also filter the stellar signal. Thus, the three filtered light curves were analyzed in parallel. In spite of the data treatment, modulations related to the {\it Kepler} year and quarters are still present in the light curves of some targets. For this reason and because we are interested in relatively slow rotators whose signal is most affected by instrumental artifacts, we obtained three additional KEPSEISMIC light curves. For these, the long-term {\it Kepler} modulation was subtracted from the light curve before the filters were applied. To do this, the autocorrelation function (ACF) of the light curve was computed and the maximum around $\pm$ 30 days of the \emph{Kepler} orbital period (372.5 days) was determined, hereafter $P_\text{Kepler}$. The original data were then phase-folded with $P_\text{Kepler}$ , and the resulting curve was fit with a third-degree polynomial. To detrend the light curve, each segment of the length $P_\text{Kepler}$ was corrected by subtracting the fitted curve linearly scaled from the data in this segment. This procedure minimized any trends around the \emph{Kepler} orbital period in the light curve for most of the targets. Nevertheless, when the retrieved $P_{\rm{rot}}$ with the new calibration was compatible with the period given in S19-21, we kept the latter for the sake of clarity. The parallel inspection of the seven light curves and respective rotation diagnostics allowed us to verify the nature of the signal. The rotation diagnostics were a wavelet analysis, an autocorrelation function, and their composite spectrum. We also visually inspected the power spectrum density of each light curve. For details about the rotation diagnostics, see \citetalias{2019ApJS..244...21S}-\citetalias{santosSurfaceRotationPhotometric2021} and references therein.\\

The results of these individual visual inspections are listed in Table \ref{tab:FineCheck} for all targets of the optimistic sample. The full version of the table along with the data of all 99 solar-type stars of this sample are available online (62 stars at solar metallicity and 37 stars with other metallicities) \footnote{\emph{Kepler} light curves and the rotation analysis we used to perform the individual visual inspections can be accessed along with a machine-readable form of Table \ref{tab:FineCheck} through the following repository: \href{https://github.com/qnoraz/AntisolarCandidatesInspections.git}{https://github.com/qnoraz/AntisolarCandidatesInspections.git}}. We discuss the examples listed in the short version of Table \ref{tab:FineCheck} below.

\begin{figure}
    \centering
    \includegraphics[width=0.8\linewidth]{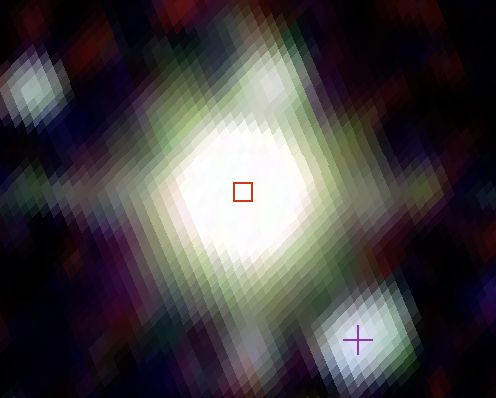}
    \caption{Image from the KASOC website of KIC~10934680 using Two Micron All Sky Survey (2MASS) images \citep{skrutskieTwoMicronAll2006}. This star is marked by a purple cross, and it is in the halo of another \textit{Kepler} target, KIC~10989776 (marked with a red square), whose \emph{Kepler} magnitude is 2.5 times brighter. }
    \label{fig:kic10934680}
\end{figure}

The star KIC~10934680 is in the halo of KIC~10989776, a 2.5 times brighter star (halo flag 1; see Fig.~\ref{fig:kic10934680}). We were unable to define the pollution clearly or clearly reject it, therefore the neighboring flag is noted 2. We therefore rejected this target from the final list of anti-solar candidates (candidate flag 0). The rotational modulation of the target is not clear either, which potentially biases the measured $P_{\rm rot}$. We therefore set the $P_{\rm rot}$ flag to 1. Moreover, if   no pollution from the neighboring star is found, we suspect that KIC~10934680 is a close multiple star system because the visible shape of this star is elongated (shape flag 1).

KIC~4042415 is also in the halo of a brighter star (halo flag 1), but this brighter star is distant and not visible in the \textit{Kepler} acquisition FoV. However, the bright star clearly pollutes the light curve of target KIC~4042415 ($P_{\rm rot}$ flag as 1), so that we rejected this target. Moreover, a different close neighbor is visible in this FoV. It is far enough and not in the photometry aperture. Thus, the neighboring flag is 1, meaning that there is a close neighbor that does not pollute the light curve.

Individual inspections of KIC~5024984 do not show any sign of pollution by the halo of a brighter star (hence the halo flag is 0). We note that a neighbor (KIC~5024961) is visible in the 40''$\times$40'' FoV around our target, but does not pollute the signal we used (the neighboring flag is 1). Nevertheless, the related uncertainties of KIC~5024984 log$_{10}$g$=4.47_{-0.75}^{+0.08}$ are too high to decipher whether it is a subgiant or an MS star. We note that the value reported for this target in the DR25 catalog \citep{mathurRevisedStellarProperties2017} is significantly different and also highly uncertain: log$_{10}$g$=3.71_{-0.27}^{+0.80}$. As we performed a strict selection, we did not consider this target in our final sample of promising candidates (hence the log$_{10}$g flag is 1).

For KIC~3860063, we find a very close neighbor in the acquisition FoV that clearly pollutes the light curve. We therefore note the neighboring flag as 3. We further suspect that the large $S_{\rm ph}$ could result from a triggered activity by tidal interactions, which could indicate that both stars interact gravitationally (see the discussion in Sect.~\ref{sec:5}).

The example KIC~6541604 shows no evidence for pollution from another star in the light curve. The new KEPSEISMIC light curves and their rotational analysis show that the $P_{\rm rot}$ extracted by \citetalias{2019ApJS..244...21S}-\citetalias{santosSurfaceRotationPhotometric2021} can be improved with the new treatment of the light curves. Hence, we are able to compute a better $P_{\rm rot}=66.0\pm11.5$ days, and then updated the Rossby number accordingly. We find $Ro_{\rm f}/{\rm Ro_{\rm f,\odot}}=1.57\pm0.29$. Its lower limit $1.28$ is below the selection threshold $Ro_{\rm f}/{\rm Ro_{\rm f,\odot}}=1.3/0.9=1.44$. Thus, the $P_{\rm rot}$ flag is 2, and we did not consider this target in the final list of candidates.

The same conclusions hold for KIC~6436380, with a new rotation value of $P_{\rm rot}=40.0\pm2.56$ days. However, when the Rossby number was accordingly updated, we find $Ro_{\rm f}/{\rm Ro_{\rm f,\odot}}=1.62\pm0.15$, which is still above the selection threshold on $Ro_{\rm f}$. Thus, we set the $P_{\rm rot}$ flag at 3 and consider KIC~6436380 as a candidate (see Table \ref{tab:ZothTargets}).

\begin{landscape}
\begin{figure}
    \centering
    \includegraphics[width=\linewidth]{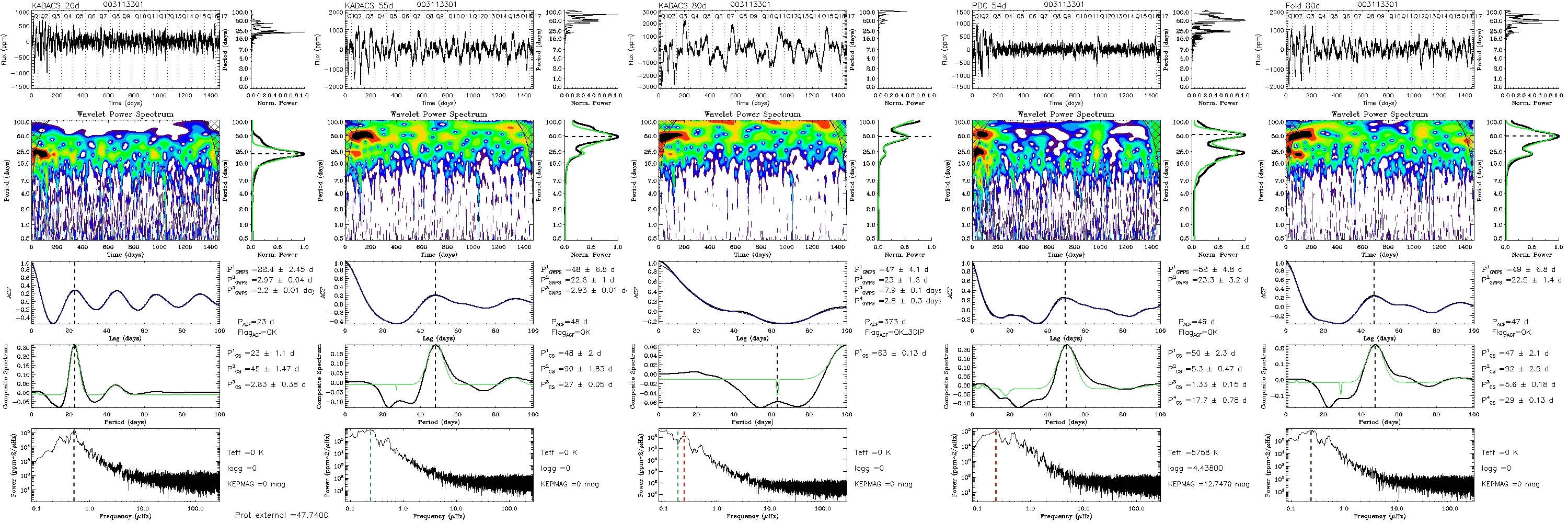}
    \caption{Example of one of the control plots for KIC~3113301. It contains the output of our rotation pipeline (see S19-21) for the three KEPSEISMIC light curves in S19-21 (filtered at 20, 55, and 80d) for the PDC-MAP light curve (which in this example  clearly filters the stellar rotation signature), and for the new KEPSEISMIC light curve filtered at 80~days in which the modulation related to $P_\text{Kepler}$ is removed. For a full description of the plots, we refer to S19-21. Briefly, from top to bottom we show the light curve and the Lomb-Scargle periodogram, the time-period analysis with the global wavelet power spectrum (GWPS), the ACF, the CS, and the full power spectrum density (PSD) in log-log scale.}
    \label{fig:kic3113301}
\end{figure}
\end{landscape}

In the final example, we did not attest any pollution or close neighboring star for KIC~3113301. This target has a circular shape in the FoV, and the amplitude of its light-curve modulation is typical of stars with similar properties. Finally, the new KEPSEISMIC light curves and their rotational analysis (see Fig.~\ref{fig:kic3113301}) confirm the reliability of its $P_{\rm rot}$ as extracted by \citetalias{2019ApJS..244...21S}-\citetalias{santosSurfaceRotationPhotometric2021}. Hence, we set all the flags to 0 except for the last flag, which indicates that we consider this target as a candidate for anti-solar DR.

\begin{table*}
  \centering
  \caption{Results of individual visual inspections.\protect\footnotemark}
  \label{tab:FineCheck}
  \begin{tabular}{lcccccc}
    \hline
    \hline\\[-2.0ex]
    KIC &  Halo  &  Neighboring &  Shape  &  log$_{10}$g  &  $P_{\rm rot}$  &  Candidate\\[0.5ex]
        &  flag  &  flag        &  flag   &  flag         &  flag           &  flag     \\[0.5ex]
        &  [0,1] &  [0,1,2,3]   &  [0,1]  &  [0,1]        &  [0,1,2,3]      &  [0,1]    \\[0.5ex]
    \hline
    \hline\\[-2.0ex]
    ... & ... & ... & ... & ... & ... & ... \\[0.5ex]
    3113301 & 0 & 0 & 0 & 0 & 0  & 1 \\[0.5ex]
    ... & ... & ... & ... & ... & ... & ... \\[0.5ex]
    3860063  & 0 & 3 & 0 & 0 & 1 & 0 \\[0.5ex]
    ... & ... & ... & ... & ... & ... & ... \\[0.5ex]
    4042415  & 1 & 1 & 0 & 0 & 1  & 0 \\[0.5ex]
    ... & ... & ... & ... & ... & ... & ... \\[0.5ex]
    5024984  & 0 & 1 & 0 & 1 & 0  & 0 \\[0.5ex]
    ... & ... & ... & ... & ... & ... & ... \\[0.5ex]
    6436380  & 0 & 1 & 0 & 0 & 3  & 1 \\[0.5ex]
    ... & ... & ... & ... & ... & ... & ... \\[0.5ex]
    6541604  & 0 & 1 & 0 & 0 & 2  & 0 \\[0.5ex]
    ... & ... & ... & ... & ... & ... & ... \\[0.5ex]
    10934680 & 1 & 2 & 1 & 0 & 1 & 0 \\[0.5ex]
    ... & ... & ... & ... & ... & ... & ... \\[0.5ex]
    \hline
    \hline
  \end{tabular}
  \tablefoot{Summary of the verifications made of the whole optimistic sample to determine whether a target is a reliable anti-solar DR candidate. The halo flag indicates whether the target is in the light halo of a brighter star from the visual inspection of 2MASS images (1) or not (0). The neighboring flag indicates whether a neighbor is at least visible in the \textit{Kepler} acquisition FoV (1, 2, and 3) or not (0). If this is the case, we determined whether the main target light curve might be (2) or is clearly (3) polluting the target star, or if we acknowledge no pollution of the rotational signal (1). The shape flag indicates whether the shape of the target in the FoV is deformed or elongated, suggesting multiple stars (1) (possibly a close companion in the FOV), or not (0). The log$_{10}$g flag indicates whether an uncertainty in the log$_{10}$g value is suspiciously large and might cause us to consider a subgiant target. When we concluded during the inspection that the $P_{\rm rot}$ value extracted by \citetalias{2019ApJS..244...21S}-\citetalias{santosSurfaceRotationPhotometric2021}  clearly derived from the rotational signal of the target (0) or not (1). If the light curve showed a rotational signal and we were able to improve the measurement and provide a new reliable $P_{\rm rot}$ , we provide flags 2 and 3. In this case, we indicate whether the new corresponding $Ro_{\rm f}/{\rm Ro_{\rm f,\odot}}$ value is high enough to keep the target above the selection threshold and in the optimistic sample (3) (i.e., if the new lower limit of this value is higher than 1.3/0.9=1.44) or not (2).
 We finally indicate with the candidate flag whether we consider the target as an anti-solar DR candidate (1) or not (0).\\ }
\end{table*}
\footnotetext{The full table is available in a machine-readable format at the public repository \href{https://github.com/qnoraz/AntisolarCandidatesInspections.git}{https://github.com/qnoraz/AntisolarCandidatesInspections.git} and at the CDS via anonymous ftp to cdsarc.u-strasbg.fr (130.79.128.5) or via http://cdsweb.u-strasbg.fr/cgi-bin/qcat?J/A+A/ .}

\newpage

\section{Scaling laws}\label{sec:CleanTargetRestrict}
\begin{figure*}
    \centering
    \includegraphics[width=0.85\linewidth]{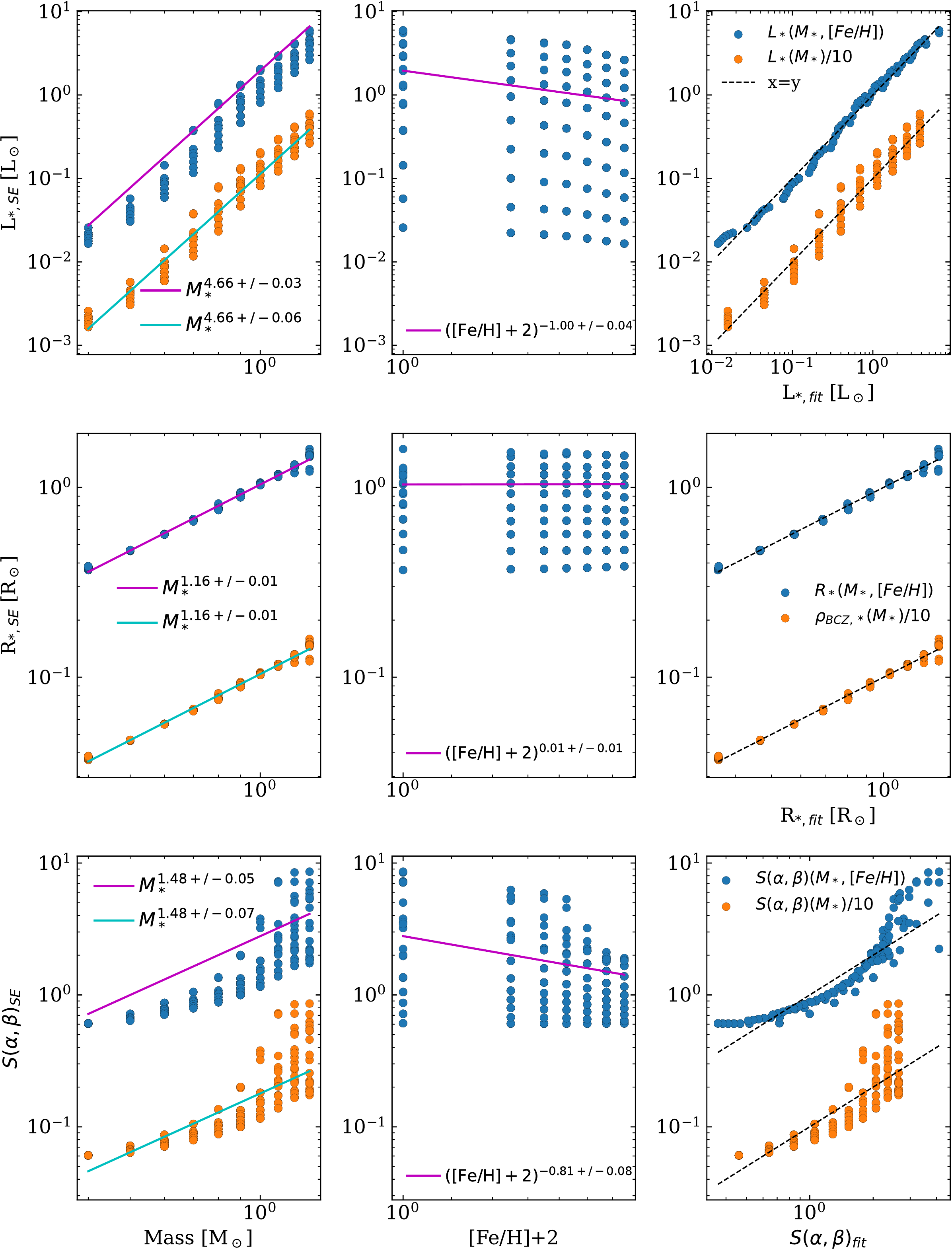}
    \caption{Linear regressions for determining the homology relations of Eqs.~\ref{eq:ScalingLawsL&R} and \ref{eq:FeHscaling} and the structure term $S(\alpha,\beta)$ as defined in Eq.~\ref{eq:RofSab} and \ref{eq:SabFeH}. We use the grid evolutionary track of \citetalias{amardFirstGridsLowmass2019} taken from the STAREVOL (SE) code. Each dot represents one of these stellar tracks, from which different values have been deduced as mean value during their MS (subscript SE). The first row shows $L_*$, the second row shows $R_*$ , and the third row shows $S(\alpha,\beta)$. The first column illustrates these different values as a function of $M_*$ , and the second column shows them as a function of $({\rm [Fe/H]}+2)$. The third column illustrates the quality of the fits. It represents the different original values ($L_*$, $R_*$, $\text{and }S(\alpha,\beta)$) as functions of the values that were estimated by the fits (subscript fit). These regressions are illustrated in the first and second columns. They are shown in purple if they are multiparametric (as a function of $M_*$ and ${\rm [Fe/H]}$) and in cyan when they are only a regression on $M_*$. Blue dots represent models on which we perform the multiparametric regression, and the orange group represents models on which we only fit on $M_*$. For the latter, we divided the values by 10 for illustration purposes.}
    \label{fig:ScalingLaws}
\end{figure*}

In order to determine the different exponents for equations \ref{eq:A19indexes} and \ref{eq:RofZAN}, we performed linear regression fits, using a least-squares method in log space on stellar evolution models from \citetalias{amardFirstGridsLowmass2019}. They proposed a grid of evolutionary tracks from the PMS to the TAMS for different initial rotation periods. Stellar masses ranged between $0.2$ and $1.5M_\odot$, and metallicities were set for 7 ${\rm [Fe/H]}$ values from $-1$ to $+0.3$. We extracted the mean value of $L_*$, $R_*$ , and $S(\alpha,\beta)$ during the MS for each model and performed regression fits as a function of $M_*$ and ${\rm [Fe/H]}$, which we illustrate in Figure \ref{fig:ScalingLaws}.

As shown in the upper panels of Fig.~\ref{fig:ScalingLaws}, multiparametric regressions give $m=m_1=4.66\pm0.03$ and $m_2=-1.00\pm0.04$. We note that the regression was improved when we added the dependence on the metallicity (purple lines). This is shown in the top right corner, where we represent the reference values as a function of fitted values. The spread of dots around the dashed function $x=y$ is smaller for the blue (multiparametric relation) than for the orange dots (only with the $M_*$ dependence). It is also visible in the uncertainties of the fit in the top left corner, which is smaller ($\pm0.03$) when we prescribe a dependence on ${\rm [Fe/H]}$. Similarly, we find $n=n_1=1.16\pm0.01$ in the second row, and we note that the quality of the regression is not improved significantly with a dependence on ${\rm [Fe/H]}$. We find that $R_*$ is not sensitive to the metallicity, with $n_2=0.01\pm0.01$. Finally, the third row shows that $q=q_1=1.48\pm0.05$ and $q_2=-0.81\pm0.08$. We note that a power law is a good approximation for the homology relations in Eqs.~\ref{eq:ScalingLawsL&R} and \ref{eq:FeHscaling}, but it is a rough approximation for the structural term $S\propto M_*^q$ when we consider a range of masses that are too high. We therefore restricted our fits for $0.4\leq M_*\leq 1.3 M_\odot$, which defines the validity range for these regressions. Moreover, we note that the multiparametric regression of $S(\alpha,\beta)$ is much better than its single-parametric regression, which is shown with the dots that are spread in the lower right corner.

\end{appendix}

\end{document}